\documentclass{aa}

\usepackage{subfig}
\usepackage{amsmath}
\usepackage{times}
\usepackage{natbib}
\usepackage[colorlinks,linkcolor=blue,anchorcolor=blue,citecolor=blue]{hyperref}
\usepackage[hyperref]{backref}
\usepackage{txfonts}
\usepackage[T1]{fontenc} 

\usepackage{graphicx}
\usepackage{color}
\usepackage{natbib}
\usepackage{rotating}
\usepackage{longtable,lscape}

\newcommand{\chandra}{{\sl Chandra}}

\newcommand{\wav}{{\tt wavdetect}}
\newcommand{\xspec}{{\it XSPEC}}
\newcommand{\kT}{{\it kT}}

\def\ltsima{$\; \buildrel < \over \sim \;$}
\def\simlt{\lower.5ex\hbox{\ltsima}}
\def\gtsima{$\; \buildrel > \over \sim \;$}
\def\simgt{\lower.5ex\hbox{\gtsima}}

\begin{document}

\title{The Swift X--ray Telescope Cluster Survey: data reduction and
  cluster catalog for the GRB fields}

\subtitle{}

\author{Elena Tundo\inst{1}, Alberto Moretti\inst{2}, Paolo
  Tozzi\inst{1}\inst{,3}, Liu Teng\inst{1}, Piero
  Rosati\inst{4}, Gianpiero Tagliaferri\inst{5}, Sergio Campana\inst{5} }

\offprints{E. Tundo, \email{tundo@oats.inaf.it}}

\institute{ $^1$INAF, Osservatorio Astronomico di Trieste, via
  G.B. Tiepolo 11, I--34131, Trieste, Italy\\ $^2$INAF, Osservatorio
  Astronomico di Brera, Via Brera 28, I--20121, Milano,
  Italy\\ $^3$INFN-- National Institute for Nuclear Physics, via
  Valerio 2, I--34127, Trieste, Italy\\ $^4$ESO, Karl-Schwarzchild
  Strasse 2, 85748 Garching, Germany\\ $^5$INAF, Osservatorio
  Astronomico di Brera, Via Bianchi 46, I--23807, Merate (LC), Italy}

\date{June 13th, 2012}


\abstract
{}
{We present a new sample of X--ray selected galaxy groups and clusters
  serendipitously observed with the X--ray Telescope (XRT) on board of
  the Swift satellite. Using the XRT archive as of April 2010, we
  searched for extended sources among 336 GRB fields with galactic
  latitude $|b|>$20$^{\circ}$. Our selection algorithm provides us
  with a flux-limited sample of 72 X--ray groups and clusters with a
  well defined selection function and an expected negligible
  contamination. The sky coverage of the survey goes from the total 40
  deg$^2$ to 1 deg$^2$ at a flux limit of $10^{-14}$ erg s$^{-1}$
  cm$^{-2}$ ($0.5-2$ keV). This paper provides a description of the
  XRT data processing, the statistical calibration of the
  survey, and the catalog of detected cluster candidates.}
{All the X--ray sources are detected in the Swift-XRT soft ($0.5-2$ keV)
  images with the algorithm {\tt wavdetect}.  A size parameter defined
  as the half power radius (HPR) measured inside a box of 45$\times$45
  arcsec, is assigned to each source. We select extended sources by
  applying a threshold on the HPR.  Thanks to extensive simulations,
  we are able to calibrate the threshold value, which depends on the
  measured net counts inside the box and on the image background, in
  order to identify all the sources with a probability $\simeq 99$\%
  of being extended. The net counts associated to each extended source
  are then computed by simple aperture photometry. }
{We compute the logN--logS of our sample, finding very good agreement
  with previous deep cluster surveys.  We did not find any correlation
  between the cluster and the GRB positions. A cross correlation with
  published X--ray catalogs shows that only 9 sources were already
  detected, none of them as extended. Therefore, $\sim 90$\% of our
  sources are new X--ray detections.  We also cross correlated our
  sources with optical catalogs, finding 20 previously identified
  clusters.  Overall, about $\sim 65$\% of our sources are new detections,
both as X--ray sources and as clusters of galaxies.}
{The XRT follow--up observation of GRBs is providing an excellent
  serendipitous survey for groups and clusters of galaxies, mainly
  thanks to the low background of XRT and its constant angular
  resolution across the field of view.  A significant fraction of the
  sample ($\sim 33$\%) has spectroscopic or photometric redshift
  thanks to a cross-correlation with public optical surveys. }

 
\authorrunning{E. Tundo et al.}

\titlerunning{The Swift X--ray Telescope Cluster Survey}

\maketitle

\section{Introduction}

X--ray observations of clusters of galaxies over a significant range
of redshifts have been used to investigate the chemical and
thermodynamical evolution of the X--ray emitting Intra Cluster Medium
\citep[ICM, see][]{2004Ettori,2007Balestra,2008Maughan,2009Anderson},
and to const|rain the cosmological parameters and the spectrum of the
primordial density fluctuations \citep{2002Rosati,2005Schuecker,
  2005Voit,2008Borgani,2009Vikhlinin,2010Mantz, 2011Allen}.  In this
respect, X--ray surveys of clusters of galaxies represent a key tool
for cosmology and the physics of large scale structure.  The need of
assembling larger and larger X--ray selected cluster samples with well
defined completeness criteria is one of the critical issues of
present-day cosmology.  In order to build statistically complete
cluster catalogs, a wide and deep coverage of the X--ray sky is
mandatory.

To date, there is a remarkable lack of recent wide area X--ray surveys
suitable to this scope.  Most of the existing cluster surveys are
based on source samples selected by ROSAT, and confirmed through
optical imaging and spectroscopic observations.  The most recent
constraints on cosmological parameters from X--ray clusters are based
on the \chandra\ follow--up of 400 deg$^2$ ROSAT serendipitous survey
and of the All-Sky Survey \citep{2009Vikhlinin,2010Mantz}.  Renewed
interest in the field of cosmological tests with clusters has been
recently provided by Sunyaev-Zel'dovich (SZ) surveys from the South
Pole Telescope Survey \citep{spt2012} and the Atacama Cosmology
Project \citep{act2012}.  At present, only modest improvements have
been obtained with respect to constraints from WMAP7 plus baryonic
acoustic oscillations plus Type Ia supernova.  The future of SZ cluster
surveys is very promising, but at present an X--ray follow--up of SZ
clusters is still needed, either for narrowing down the uncertainties
on the cluster mass, or to firmly evaluate purity and completeness of
the sample.  For example, a large effort is being invested in the
X--ray follow--up with \chandra\ (PI B. Benson) of SZ selected clusters
from the South Pole Telescope Survey. This follow--up will provide the
X--ray data for 80 massive clusters spread over 2000 deg$^2$ in the
redshift range 0.4$<z<$1.2.

\begin{center}
\begin{table*}
\centering
\caption{Flux limited X--ray Cluster Surveys}
\label{surveys}
\begin{tabular}{@{}lcccc@{}}\hline
\smallskip
Name   & Flux limit       & solid angle & Number of sources & Reference \\ 
       & cgs ($0.5-2$ keV)  &    deg$^2$   &                 &            \\ 
\hline 
\hline 
\smallskip

SEXCLAS & $0.6 \times 10^{-14}$  (min) & 2.1  &  19 & Kolokotronis et al. (2006)\\
DCS & $0.6\times 10^{-14}$  (min) & 5.55  & 36 & Boschin (2002) \\
ChaMP & $1.0\times 10^{-14}$   (min)& 13.0  & 49 & Barkhause et al. (2006) \\
\bf SXCS  & \bf $1.0\times 10^{-14}$   (min)  & \bf 40.0  &\bf 72  & \bf This work \\
XDCP & $1.0 \times 10^{-14}$  (average)  & 76.0  & 22 ($z>0.9$) & Fassbender et al. (2011)  \\
XCLASS & $2 \times 10^{-14}$ (min) & 90.0 & 347 & Clerc et al. (2012)\\
Peterson09 &  $ \sim 0.3\times 10^{-14}$ (min) & 163.4& 462 & Peterson et al. (2009)\\
XCS  &  $>300$ net cts  & 410.0  &  993  & Lloyd-Davies et al. (2011)\\
\hline 
\hline 
\smallskip

SXDF & $0.2 \times 10^{-14}$  (min) & 1.3  & 57 & Finoguenov et al. (2010)\\
COSMOS  & $0.2 \times 10^{-14}$  (min) & 2.1 & 72  & Finoguenov et al. (2007)\\
XMM-BCS & $0.6 \times 10^{-14}$   (min)     & 6.0 & 46  & Suhada et al. (2012)\\
XMM-LSS & $\sim  10^{-14}$   (min)     & 11.0 & 66  & Adami et al. (2011)\\
\hline 
\hline 
\smallskip
\end{tabular}
\tablefoot{List of X--ray flux limited cluster surveys based on
  \chandra\ or XMM data, updated at the time of writing (May 2012),
  plus the Swift/XRT Cluster Survey presented in this work
  (highlighted in bold).  Surveys based on archival data are listed in
  the upper part of the table, while dedicated (contiguous) surveys
  are shown in the lower part.  Surveys are ranked according to the
  total solid angle.  The total number of clusters refer to the X--ray
  selected only, while the quoted solid angle is the maximum covered
  by the survey.  Note that the limiting soft fluxes quoted in this
  table (when available) may refer to the lowest value in the sample
  (minimum) or to an average value over the entire solid angle
  (average).  For a full characterization of the survey depth, i.e.,
  the sky coverage as a function of the flux, we refer to the
  corresponding papers.}
\end{table*}
\end{center}

New X--ray surveys of clusters of galaxies in the \chandra\ and
XMM--Newton era are based on the compilation of serendipitous medium
and deep--exposure extragalactic pointings not associated to
previously known X--ray clusters.  Among these, one of the first
survey was assembled by \citet{2002Boschin}, who found 36 clusters
(among them 28 new detections) in 5.55 deg$^2$ of surveyed area.
Eventually, the \chandra\ Multiwavelength Project (ChaMP)
Serendipitous Galaxy Cluster Survey \citep{2006champs} identified
about 50 cluster and group candidates from 130 archival
\chandra\ pointings covering 13 deg$^2$. Of the 50 clusters, about 16
are expected to have redshift $z>$0.5. 

More effort is devoted to the XMM-Newton archival data, also motivated
by the larger solid angle and the nominal larger sensitivity.  The
XMM--Newton Distant Cluster Project \citep[XDCP,][]{2011fassbender}
take advantage of an optical and IR follow--up of extended source
candidate in XMM images, to identify high--$z$ clusters.  So far, the
XDCP survey has yielded 22 spectroscopically confirmed clusters in the
redshift range 0.9$<z\simlt$1.6.  A key step in XDCP is the
identification of high--$z$ candidates based on optical and NIR
photometric technique,as well as extensive spectroscopic work. In this
respect, it currently provides the largest sample of confirmed galaxy
clusters at $z>$0.8, and the purity of the sample is extremely high.
Another small survey (2.1 deg$^2$) conducted using XMM archival data
is SEXCLAS \citep{kolo2006}, which include 19 serendipitous detections
down to $6\times 10^{-15}$ erg s$^{-1}$ cm$^{-2}$.

The largest project based on the entire XMM archive is the XMM
Clusters Survey \citep[XCS,][]{romer01,2011lloyd-davies}, with about
1000 cluster candidates over a solid angle of 410 deg$^2$ (the largest
based on current X--ray facilities).  Recently, about 500 clusters
have been optically identified out of $\sim 1000$ candidates
\citep{2012Mehrtens}.  Another survey based on the XMM archive is
XCLASS, which is limited to brighter sources and aims at constraining
cosmological parameters on the basis of the X--ray information only
\citep{clerc12}.  Finally, a survey combining 41.2 deg$^2$ of XMM and
\chandra\ overlapping archival data, plus 122.2 deg$^2$ of
\chandra\ only, has been presented in \citet{peterson09}, for a total
of 462 new serendipitous sources. Clearly, there is a significant
overlap among the cluster samples derived from serendipitous
XMM/\chandra\ surveys.

In addition, there are also ongoing dedicated, contiguous surveys. The
cluster survey in the Subaru-XMM Deep Field (SXDF) reaches a depth of
$2\times 10^{-15}$ erg s$^{-1}$ cm$^{-2}$ over 1.3 deg$^2$ with 57
X--ray clusters identified with the red-sequence technique
\citep{finog2010}.  Similar depth has been reached in the COSMOS field
over 2 deg$^2$, for a total of 72 clusters \citep{2007Finoguenov}.
The XMM-Newton-Blanco Cosmology Survey project \citep[XMM-BCS,
][]{suhada2012} is a multiwavelength X--ray, optical and mid-infrared
cluster survey covered also by the South Pole Telescope and the
Atacama Cosmology Telescope with the aim of studying the cluster
population in a 14 deg$^2$ field.  The analysis of the first 6 deg$^2$
provided a sample of 46 clusters.  Finally, the largest contiguous
survey is the XMM Large Scale Structure Survey
\citep[XMM-LSS,][]{2007XMMLSS} which is covering a region of 11
deg$^2$ with the aim of tracing the large scale structure of the
Universe out to a redshift of $z\sim$1. At present, a first data
release from an area of 6 deg$^2$ consists in 66 spectroscopically
confirmed clusters with 0.05$<z<$1.5 \citep{2011Adami}.

One of the most important goal of these surveys is to find massive
clusters at high $z$.  Clearly, tracing the most massive,
gravitationally bound structures in the Universe up to the highest
possible redshift is of paramount importance for structure formation
and for cosmology.  While the redshift limit for X--ray selected
clusters is $z\sim 1.57$ (in XDCP), extended X--ray emission has been
detected in optical and IR selected clusters at $z=1.75$
\citep{2012Stanford}, $z=2.07$ \citep{2011Gobat}, with an extreme
candidate at $z\sim 2.2$ \citep[see ][]{2011andreon}.

The present situation is summarized in Table \ref{surveys}.  This
picture is not expected to change significantly in the next years,
with a modest increase in the source statistics and in the quality of
the X--ray data.  In particular, the study of distant ($z\geq 1$)
clusters relies almost exclusively on time-expensive follow--up with
\chandra.  In the case of \chandra , the process of assembling a wide
and deep survey is slow due to the small field of view (FOV) and the
low collecting area.  In the case of XMM-Newton the collecting area is
significantly higher, but the identification of extended sources, in
particular at medium and high redshift, is more difficult due to the
larger size of the Point Spread Function (PSF, whose half energy width
is 15" at the aimpoint) and its degradation as a function of the
off--axis angle.  This is not an issue for nearby or medium redshift
clusters and groups, but it becomes a problem at $z>1$, where the
optical and IR data play a dominat role in cluster identifications.
In addition, the relatively high and unstable background may hamper
the proper characterization of low surface brightness sources.  In
conclusion, while their design is optimized to obtain detailed images
of isolated sources to explore the deep X--ray sky, a substantially
different mission strategy is required for surveys.

\begin{figure}
\centering
\includegraphics[width=\columnwidth]{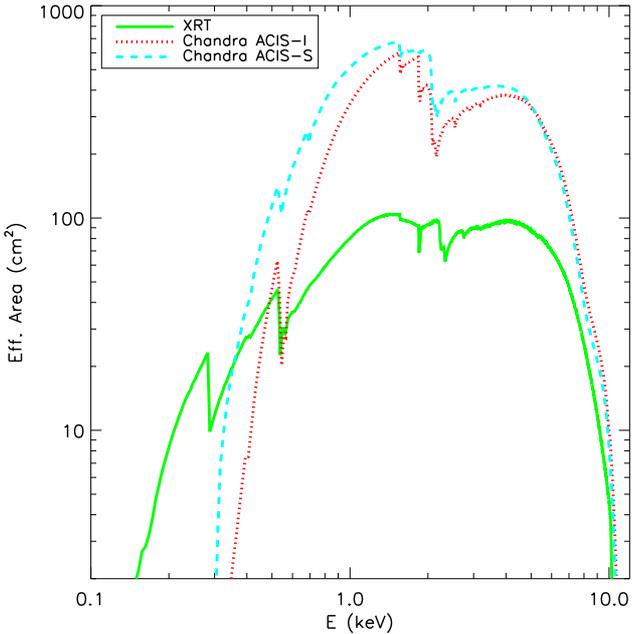}
\caption{Effective area of XRT (green solid line) as a function of the
  energy compared with that of the \chandra\ satellite (red dotted and
  cyan dashed lines for \chandra\ ACIS-S and ACIS-I, respectively).}
\label{xrt_arf}
\end{figure}

In this work we present a new X--ray cluster survey using the rich
archive of the X--ray Telescope (XRT) onboard of the Swift satellite
\citep{2005bur} which has never been used for the detection of
extended sources.  Despite its low collecting area (about one fifth of
that of \chandra\ at 1 keV, see Figure \ref{xrt_arf}), XRT has two
characteristics which are optimal for X--ray cluster surveys: a low
background and an almost constant PSF across the FOV
\citep{2007Moretti}.  Moreover, almost all the Swift/XRT pointings can
be used to build a serendipitous survey.  In fact, the operation mode
of the XRT observations considered in this work, consists of a prompt
follow--up of fields centered on Gamma Ray Bursts (GRB) detected by
Swift.  As we will show in \S~\ref{cluster_cat}, the GRBs do not show
any correlation with our extended sources.  Therefore, despite its
small size, the XRT can be successfully used to build an unbiased
cluster survey.

The catalog of extended sources presented in this paper constitutes
the Swift X--ray Cluster Survey (SXCS) and it is based on the 336 GRB
fields with galaxtic latitude $|b|>$20$^\circ$ present in the XRT
archive as of April 2010.  The sources are identified thanks to a very
simple but effective algorithm to select extended X--ray sources in
the XRT soft band images.  In this paper we adopt a conservative
detection threshold which guarantees very low contamination and a well
defined completeness function.  This catalog is complementary to the
catalog of point sources identified in the GRB fields of XRT by
\citet{2011Puccetti}, which focus on the study of the AGN population.

The paper is organized as follows. In \S~\ref{sur_sec} we provide a
description of the principal characteristics of the XRT. In
\S~\ref{field_sel} we describe the field selection and the data
reduction.  In \S~\ref{ext_sel} we describe the detection algorithm and the
selection of the extended source candidates.  In \S~\ref{ext_phot} we
describe how we performed extended source photometry. In \S~\ref{cluster_cat}
we present the final list of groups and clusters in our sample,
compute the sky coverage and the LogN--LogS, and check for possible
selection bias.  In \S~\ref{cross_corr} we correlate our catalog with existing
databases to identify previously known sources and collect the
spectroscopic or photometric redshifts of member galaxies available in
the literature. In \S~\ref{discus} we discuss our results in the context
of current and future X--ray surveys.  Finally, in \S~\ref{conc} we
summarize our conclusions.

\section{XRT characteristics}\label{sur_sec}

\begin{figure}
\centering
\includegraphics[width=\columnwidth]{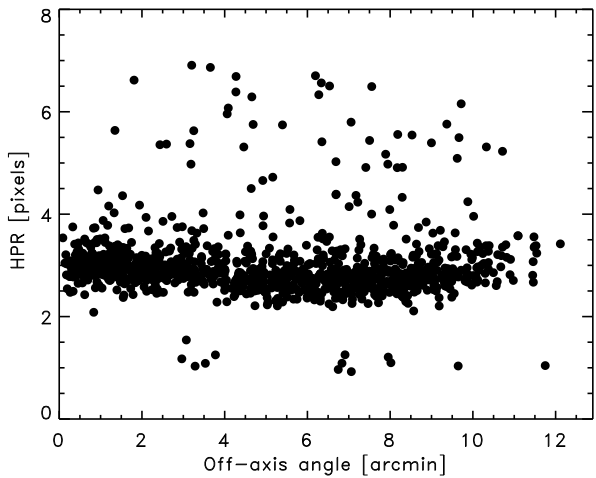}
\caption{Measured HPR within a box of $45\times 45$ arcsec as a
  function of the off-axis angle for all the sources detected in the
  XRT fields used in this work.}
\label{psf}
\end{figure}

The XRT is part of the scientific payload of the Swift satellite
\citep{2004gehrels}, a mission dedicated to the study of gamma-ray
bursts (GRBs) and their afterglows operating since January 2005\footnote{In
2012 the NASA Senior Review committee allocated full funding for for
Swift in the period 2013-14 and recommended full funding also for
2015-16 with next review in 2014.}.  GRBs are detected and localized
by the Burst Alert Telescope \citep[BAT, ][]{2005bart}, in the 15-300
keV energy band and followed--up at X--ray energies (0.3--10 keV) by
the X--ray Telescope.  The XRT \citep{2005bur} is an X--ray CCD
imaging spectrometer which utilizes the third flight mirror module
originally developed for the JET-X program \citep{1994cit}.  The
mirror module focuses X--rays (0.2--10 keV) onto a XMM-Newton/EPICMOS
CCD detector consisting of $600 \times 600$ pixels, with a nominal
plate scale of 2.36 arcseconds per pixel, which provides an effective
FOV of the system of $\sim 24$ arcmin. The PSF, similar to XMM-Newton,
is characterized by an half energy width (HEW) of $\sim 18$ arcsec at
1.5 keV \citet{2007Moretti}.  The PSF dependence on the off-axis angle
is very weak.  This is due to the fact that the CCD is intentionally
slightly offset along the optical axis from the best on-axis focus in
order to have a uniform PSF over a large fraction of the FOV, with a
negligible dependence on the photon energy. To show the remarkably
flat behaviour of the PSF, we show in Figure \ref{psf} the measured
Half Power Radius (HPR) within a box of $45\times 45$ arcsec as a
function of the off-axis angle for all the sources detected in the XRT
fields used in this work. The PSF can be analytically described by a
King function with slope $\beta \sim 1.45$ and core radius $r_c\sim
5.3$ arcsec \citep{2007Moretti}.

One of the most relevant aspect for our purposes is the low level and
the high reproducibility of the background not associated to
astronomical sources (NXB).  Due to the low orbit and short focal
length, the NXB is the lowest among the currently operating X--ray
telescopes \citep[see][]{2007hall}.  As a reference, the NXB of XRT
calculated per solid angle and normalized by the effective area is a
factor $\sim 7$ lower than \chandra\ in the 0.5-7.0 keV energy
band (Moretti et al. 2012 in preparation).

\section{Fields selection and data reduction}\label{field_sel}

\begin{figure}
\centering
\includegraphics[width=\columnwidth]{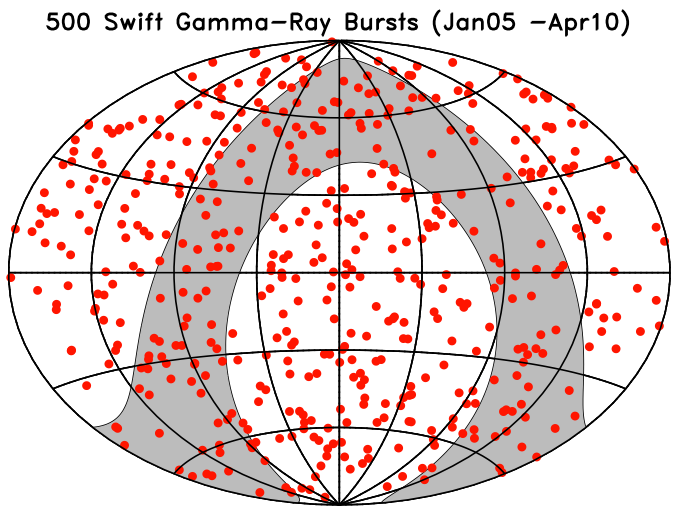}
\caption{Distribution on the sky (Aitoff projection) of the GRB fields
  in the XRT archive as of April 2010 (red dots).  The fields selected
  in our survey are those outside the Galactic plane (shaded area).}
\label{swwsky}
\end{figure}

We consider the entire Swift/XRT archive from February 2005 to April
2010, including 502 fields in the corresponding GRB positions,
distributed in the sky as shown in Figure \ref{swwsky}.  We select the
fields with Galactic latitude $|b|>$20$^\circ$, to avoid crowded
fields and strong Galactic absorption.  This selection provides us
with 336 fields which we use to search for extended sources.

In this paper we use photon counting mode data with the standard
grade selection (grade 0-12) reduced by means of the {\tt
  xrtpipeline} task of the current release of the HEADAS software
(version v6.8) with the most updated calibration at the time of
writing (CALDB version 20111031, Nov 2011).  More detail on the
standard data reduction can be found in the instrument user
guide\footnote{http://heasarc.nasa.gov/docs/swift/analysis/documentation}.
Then, we proceed with a customized data reduction, aimed at optimizing
our data to the detection of extended sources.  First we exclude the
external CCD columns (corresponding to the detector coordinates Detx
$<$ 90 and Detx $>$ 510) which are affected by the presence of
out-of-time-events from corner calibration sources \citep[see][for a
  detailed map of the CCD and a discussion on the XRT
  background]{Moretti09}.  The corrected FOV, after removal of the
external CCD columns, is 16.5$\arcmin~\times~$18.9$ \arcmin$ (0.087
deg$^2$) for each pointing.  Different observations of the same object
and corresponding exposure maps are merged by means of the the
\texttt{extractor} and \texttt{farith} tasks of the HEADAS software,
respectively.

In order to further reduce the background, we investigate the
background light curves in the soft image of each field. At the
beginning of each observation, corresponding to the maximum emission
from the GRB, the background on the entire image is significantly
larger than the typical, quiescent value.  Actually it was previosly
known \citep{Moretti09} that for very high fluences\footnote{The
  fluence is the total energy delivered per unit area, obtained by
  integrating the source flux over time.}, about 5\% of the GRB X--ray
emission is scattered across the image.

\begin{figure}
\centering
\includegraphics[width=\columnwidth]{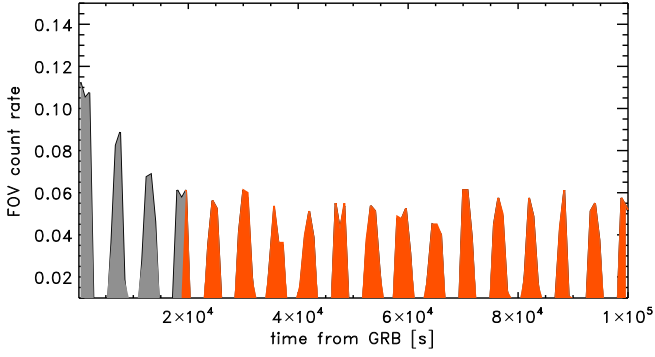}
\caption{Light curve of the background of the field of GRB051008
  plotted as a function of the time elapsed from the GRB trigger.  The
  saw-tooth appearance is due to the orbits of Swift (one orbit lasts
  90 minutes).  The removed time interval below the epoch $T_0 + 0.1
  \times t_{exp}$ is shown in grey.}
\label{light_curve}
\end{figure}

\begin{figure}
\centering
\includegraphics[width=\columnwidth]{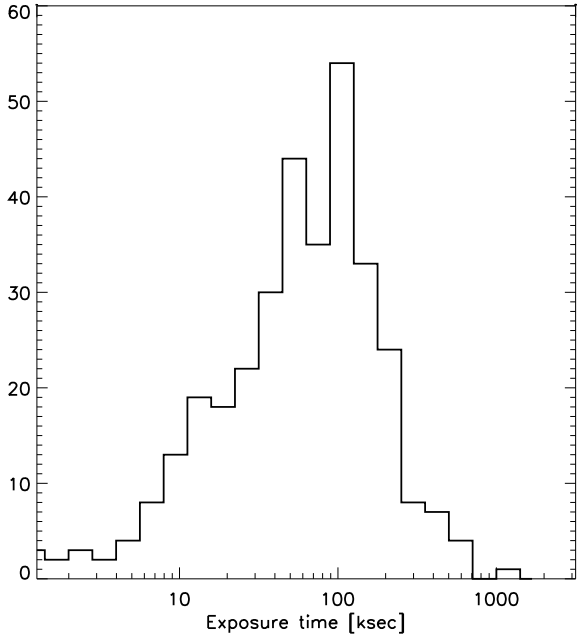}
\caption{Distribution of the exposure times in ks for the 336
  selected GRB fields.}
\label{exp}
\end{figure}

We effectively tested that by removing all the data taken before the
epoch $T_0 + 0.1 \times t_{exp}$, where $T_0$ is the epoch of the
beginning of the observation, and $t_{exp}$ is the total effective
exposure time, the average background on a typical image can be
reduced by $\sim 5$\%. This empiric rule is motivated by the fact that
the total exposure time of a typical GRB follow--up observation is set
by the GRB emission itself, therefore brighter GRBs will have larger
observation length and therefore larger time intervals removed.  In
addition, note that the time interval $0.1 \times t_{exp}$ does not
correspond to 10\% of the exposure time, since this is distributed
over several orbits, each one with an observability time of 1500 s
over a total orbit period of 5400 s.  Figure~\ref{light_curve} shows
an example of a typical light curve, where the time interval that has
been removed is shown in grey.  The removal of the data up to the time
$T_0 + 0.1 \times t_{exp}$ corresponds typically to the first two or
three orbits.  On average, the time interval $0.1 \times t_{exp}$
corresponds to 2-3\% of the effective exposure time, since this is
distributed over several days with 10-20 ks observed every day.  We
also note that removing the first orbits has the effect of reducing
the non--homogeneity of the background caused by the GRB emission,
which would have affected the detection of extended sources.  In
Figure~\ref{exp} we show the distribution of the effective exposure
times of the XRT fields after this correction.

\section{Selection of extended-source candidates}\label{ext_sel}

\subsection{Initial source detection}

The identification of the X--ray sources is performed with the
standard algorithm \wav\ within the CIAO software used successfully on
\chandra\ and XMM-Newton images \citep{2011lloyd-davies}.  We run the
algorithm on the images obtained in the soft ($0.5-2$ keV) band using
a set of six wavelet scales corresponding to 2, 4, 8, 16 and 32 image
pixels.  We use a mild selection threshold corresponding to the
\wav\ threshold parameter $10^{-6}$.  Despite not guaranteeing a high
completeness nor high purity, this choice is motivated by the fact
that the selection of extended sources will be based on the growth
curve method, and that only relatively high S/N sources will be
selected as we will show in Section 4.5.  Therefore the
\wav\ threshold parameter has no effects on our final cluster
candidates list.

The soft band is optimal to identify extended ICM emission, since in
this band XRT has the highest effective area and the lowest
instrumental background.  In this way we maximise our sensitivity in
the detection of extended source powered by thermal bremsstrahlung,
whose soft band emission has a small K-correction up to $z\sim 1$, at
least for hot clusters (\kT$>$3 keV).  We also considered modifications
to the standard soft band in order to find the energy range optimal
for cluster detection, similarly to what has been done for
\chandra\ and XMM--Newton \citep{scharf2002}.  We explored the use of
a narrower energy band in order to reduce the effect of the Galactic
background, which rapidly grows below 1 keV.  We found that a narrower
energy band would have a minor positive effect on the detection of
clusters, while negatively affecting the detection of low temperature
groups whose emission is below 2 keV.  We also explored the use of a
larger upper energy value, but the poor photons statistics for thermal
spectra at high energies and the rapidly increasing hard background
nullify any advantage for values above 2 keV.  Since the small,
positive effects we found in changing the energy band are also
significantly dependent on the temperature of the ICM, we decide to
maintain the standard $0.5-2$ keV band.

\begin{figure}
\centering
\includegraphics[width=\columnwidth]{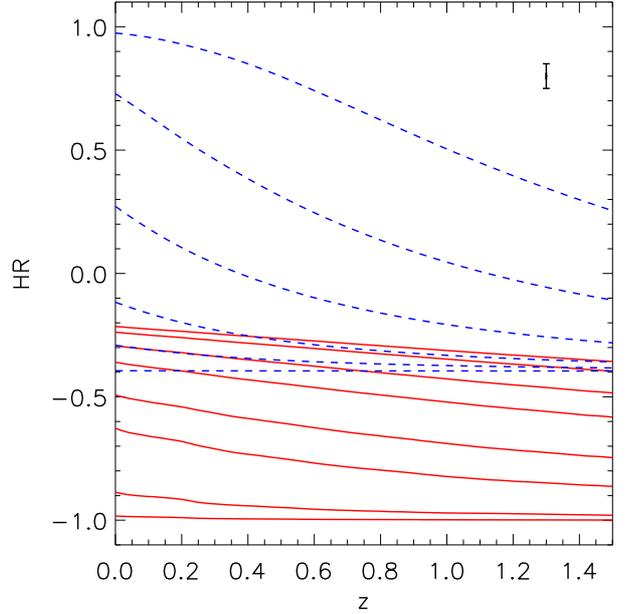}
\caption{Hardness ratios modeled for XRT as a function of redshift for
  groups and clusters with several temperatures (red continuous lines,
  $kT$ respectively equal to 0.5, 1.0, 2.0, 3.0, 5.0, 7.0, 10.0 and
  12.0 keV from the bottom to the top) and for AGN with different
  intrinsic absorption (blue dashed lines, N$_H$ respectively equal to
  0.01, 0.1, 0.3, 1.0, 3.0 and 10.0$\times$10$^{22}$ cm$^{-2}$ from
  the bottom to the top). In the top right, we show the typical error
  bar on HR for a source with 200 total net counts.  }
\label{hr_vs_z}
\end{figure}

Incidentally, we notice that, given the sensitivity at high energies
of the XRT, it is possible in principle to use also the hard band (2-7
keV) to disentangle thermal emission from power-law emission typical
of Active Galactic Nuclei (AGN), on the basis of the source hardness
ratio.  The hardness ratio is defined as HR=(H-S)/(H+S), where H and S
are the hard band and soft band counts respectively, corrected for
vignetting.  In Figure \ref{hr_vs_z} we show the hardness ratio as a
function of redshift expected for groups and clusters at different
temperatures, compared to the hardness ratio of AGN with different
intrinsic absorption N$_H$. Values for clusters and AGN are taken
using {\tt Xspec} v12.6.0 assuming in the first case an absorbed
thermal model with a mean Galactic absorption of
$N_H=3 \times 10 ^{20}$ cm$^{-2}$, and in the second case assuming an
absorbed power law with an intrinsic redshifted absorption.  We find
that only nearby, strongly absorbed AGN have hardness ratio values
clearly different from those of the ICM.  We conclude that the hard
band does not provide a relevant information for the detection of
groups and clusters.  Therefore we will use only the soft band in this
work.  The full band will be used for the spectral analysis of the
brightest sources in a companion paper (Moretti et al. in
preparation).

Since the algorithm \wav\ has never been tested thoroughly on XRT
images, we cannot immediately associate an expected contamination
level to the threshold value of $10^{-6}$.  In any case, at this stage we
are mostly interested in collecting all the possible extended-source
candidates, while the contamination level will be estimated during our
selection process.  This step leads us to identify a total of $\sim
10^4$ sources in the 336 GRB fields.  This number is somewhat lower
than that expected on the basis of the Swift serendipitous survey
\citep[SwiftFT, see][]{2011Puccetti} which was based on a $\sim 20$\%
smaller field selection.  The lowest number of net photons in our
total source list is as low as 5.  Despite the low background, Poisson
fluctuations may appear as spurious sources at such a low signal.
However, we do not apply any filter to this parent list, since much
more stringent thresholds will be applied when selecting extended
sources, as described in the next subsection.

\subsection{Source extent determination}

For each source we consider a 60$\times$60 pixels box centered around
the source.  In each of these boxes, we mask the other sources whose
emission can overlap with the central one, by removing a PSF image.
The PSF is accurately fitted with a King profile normalized to each
source photometry assuming a core radius of $r_c = 5.3$ arcsec and a
$\beta = 1.45$.  These values are valid at energies $\sim 1.5 $ keV
and have been recomputed in--flight by fitting the many high S/N
sources observed by XRT, therefore updating the values in
\citet{2005Moretti} and \citet{2007Moretti}.  This gives us a
sub--image cleaned from point-like sources, with the considered source
at its center.  In order to account for defects due to missing columns
or removed pixels in the CCD, we divide this image by the
corresponding soft exposure map.  Then, we compute the growth curve of
the source within a box of 45$\times$45 arcsec.  This region
corresponds to 2.5 times the HEW (which is 18 arcsec) and includes
80\% of the flux of a point source.  From the growth curve within this
region, we measure the HPR, effectively defined as the 50\% encircled
energy radius within a box of $45\times 45$ arcsec. Note that the HPR
is smaller than the HEW, since it refers to only 80\% of the total
flux.  The choice of a restricted regions is motivated by the need of
sampling the growth curve with a good signal-to-noise ratio (S/N) for
the majority of the sources in the initial list.  This choice does not
affect by any means the final total flux, which will be computed
according to a different procedure as shown in the next subsections,
and it is used only to classify extended versus unresolved sources.

\subsection{Point source simulation, and determination of threshold parameters}

Extensive simulations spanning all the parameter range found in the
survey (in particular source fluxes and background levels) have been
used to estimate the expected distribution of the HPR for point
sources.  The input flux distribution for unresolved sources is
consistent with the number counts obtained by \citet{2011Puccetti} and
with the deeper CDFS counts \citep{CDFS2002}, while the fluxes for
clusters are distributed according with the number counts obtained in
the ROSAT Deep Cluster Survey \citet{2002Rosati}.

\begin{figure}
\centering
\includegraphics[width=\columnwidth]{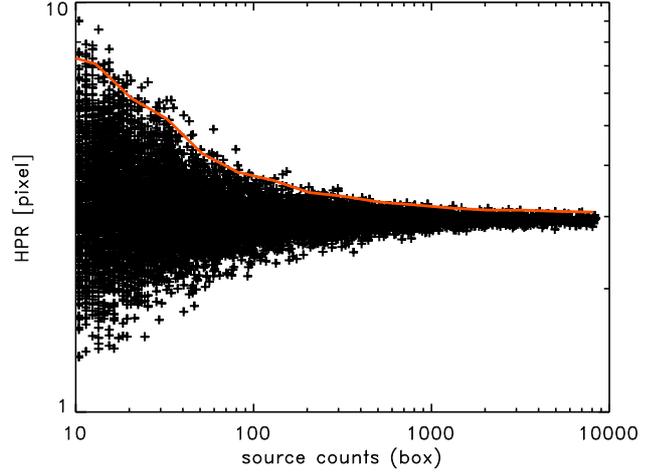}
\caption{Distribution of the measured HPR within a 45$\times$45 arcsec
  box for simulated unresolved sources as a function of the soft
  counts within the box.  The same range of exposure times and
  backgrounds present in the real data has been adopted in the
  simulations.  The continuous line shows the 99\% confidence level
  envelope of the source distribution.}
\label{simulations}
\end{figure}

The distribution of the measured HPR in the simulations as a function
of the source counts within the 45$\times$45 arcsec box is shown in
Figure \ref{simulations}. The simulations (including only point
sources) are used to define the threshold values HPR$_{th}$ above
which a source is inconsistent with being unresolved at the 99\% level
(see red line in Figure \ref{simulations}).  This criterion is
extremely simple and it does not depend on the off--axis angle
$\theta$, thanks to the flat PSF.  However, HPR$_{th}$ is a function
of the source counts measured within the 45$\times$45 arcsec box, in
particular it increases significantly below 60 counts.  Above this
value, HPR$_{th}$ ranges between 3 and 4 image pixels (equal to
7.1-9.4 arcsec for a pixel size of 2.36 arcsec).  Instead of using a
single HPR$_{th}$ based on the entire simulation, we compute
HPR$_{th}$ for a set of different background values, finding a
significant dependence which will be taken into account in the source
selection.  By directly applying this criterion, we expect to have a
number of spurious sources equal to 1\% of the total source number.
We remark that the counts within the 45$\times$45 arcsec box can be
much lower than the total net counts, particularly for extended
sources.

\begin{figure}
\centering
\includegraphics[width=\columnwidth]{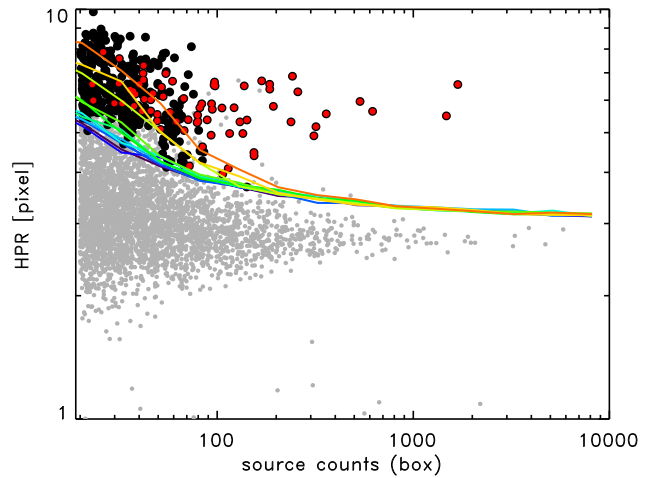}
\caption{Distribution of the measured HPR within a box of $45\times
  45$ arcsec for all the sources detected in the XRT fields as a
  function of the soft counts measured within the HPR.  All possible
  extended source candidates are marked with black big dots, while red
  big dots indicate sources with $>$100 soft net counts from the
  aperture photometry in the extraction radius (see
  Section~\ref{ext_phot}) and are therefore included in the cluster
  candidate list.  The continuous lines show the typical 99\%
  confidence level envelope for different exposures, from 1~ks (orange
  line) to 1~Ms (blue line).}
\label{swwclu1}
\end{figure}

This procedure is applied to all sources detected with \wav.  Figure
\ref{swwclu1} shows the distribution of the HPR of the detected point
sources (grey dots) versus the source counts measured inside the HPR,
with the extended source candidates marked as black and red big dots.
Note that extended source candidates are selected on the basis of a
HPR$_{th}$ value which depends also on the background (tied to the
exposure of the image).  The HPR$_{th}$ for different exposures are
shown in Figure~\ref{swwclu1} as solid lines ranging from 1~ks (orange
line) to 1~Ms (blue line).  The number of extended source candidates
grows rapidly when the number of detected counts within the box
decreases towards low values.  The inclusion of all the candidates in
our survey clearly would imply a large number of spurious sources,
given the large total number of sources ($\sim 10^4$).  We finally
apply the threshold on the total number of counts to be $>100$ in the
extraction radius (defined as the circular region where the source
surface brightness is larger than the background level).  The red big
dots indicate the candidate extended sources matching all our
selection thresholds.

The sharp threshold on the soft net counts allows us to compute
directly a sky coverage as a function of the source flux within the
extraction region (see Section~\ref{cluster_cat}).  Clearly, the sky
coverage information would be sufficient to compute the logN-logS only
in absence of any morphology bias.  Actually, we do miss a fraction of
extended sources, mostly due to the wide range of intrinsic
morphologies.  The completeness of our survey is then estimated as
described in the next subsection.

\subsection{Completeness and contamination level}

To properly characterize the quality of our sample we need to assess
the completeness, defined as the fraction of extended sources actually
selected as extended by our procedure as a function of their soft net
counts. In order to estimate the completeness, we realize another set
of simulations with a different strategy. We simulated a few thousand
extended sources, spanning a wide range of fluxes consistent with the
number counts measured in the ROSAT Deep Cluster Survey
\citep{1998Rosati}.

\begin{figure}
\centering
\includegraphics[width=\columnwidth]{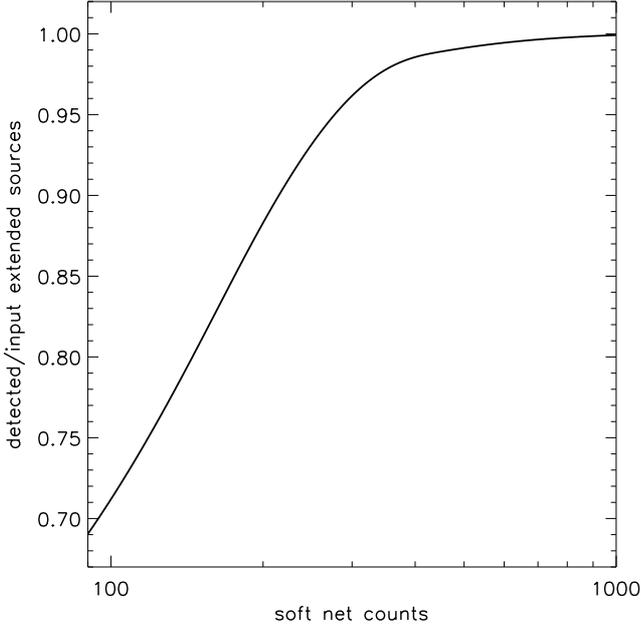}
\caption{Completeness level, expressed as the ratio of recovered
  extended sources on the number of simulated extended sources as a
  function of the input soft counts.}
\label{completeness}
\end{figure}

Our extended sources are modelled starting from ten real cluster
images originally obtained with the \chandra\ satellite (therefore
with a resolution much higher that that of XRT), cloned at a typical
redshift and resampled at the XRT resolution. This cloning procedure,
already used to investigate the evolution of cool core clusters at
high redshift \citep[see][]{2008Santos, 2010Santos} allows us to
measure our completeness for a realistic range of surface brightness
profiles, representing the local population of groups and clusters of
galaxies: temperatures go from 2 keV to 8 keV, the redshift of the
clusters are all below 0.2, and half of the clusters have a cool core.
Clearly, the limited set of group/cluster templates, and the
assumption of no evolution in the surface brightness properties with
redshift, may introduce some difference with respect to the actual
cluster population.  The impact of the intrinsic morphologies on the
source selection is properly taken into account as long as the
morphologies of the simulated sources are representative of the real
sources.  This is admittedly a limitation of the present approach.
Still, this procedure is the most accurate with the present knowledge.
The simulated clusters are positioned randomly on real XRT images and
the fluxes are converted into count rate using the response function
of the instrument, including vignetting effects. We run our detection
procedure on the simulated images, and compute the completeness as the
recovered source fraction (i.e., the fraction of sources detected and
characterized as extended with our criterion) as a function of the
input counts.  As shown in Figure~\ref{completeness}, we reach a
completeness level $\sim$70\% for sources above our threshold of
$\sim$100 net counts within the extraction regions. This level
increases to $\sim$90\% for sources with $\sim$200 net counts, and
reach a completeness level of $>$95\% for sources with $\sim$300 net
counts.  We will use this information to correct the logN-logS of our
sample in \S~\ref{cluster_cat}.  We remark that this completeness
function is by no means general: it depends, in fact, on the intrinsic
properties of the survey, on the exposure time distribution, and on
the source selection and classification method.  We are actually
aiming at increasing the completeness down to a threshold of $\sim$50
net counts, while maintaining a low contamination level, thanks to a
different detection algorithm based on Voronoi tessellation (Liu et
al. in preparation).

Finally, our simulations allow us to estimate also the contamination
level.  This is obtained directly by averaging the number of spurious
sources surviving our selection criteria in all our survey
realizations.  The average expected number of spurious sources in the
SXCS survey turns out to be $\sim 5$.  The contamination is due to Poisson
noise in the measured parameters (HPR, photometry) and to blending.
We neglect any spatial correlation which, however, is expected to give
a negligible enhancement to the contamination level.

\subsection{Final sample selection}

Our algorithm identified 87 candidate extended sources with $\geq$100
soft net counts.  We proceeded to a careful visual inspection of these
candidates, including a look through the optical images taken from the
DSS and from the SDSS when available.  This visual inspection allows
us to remove 3 sources which are identified with stars\footnote{Bright
  stars appear as extended in the XRT image due to the saturation of
  the brightest pixels. Despite this, it is easy to remove these
  spurious extended sources after a crosscorrelation with any optical
  database.}.  In addition, 3 sources are removed since they are
identified with nearby large galaxies.  Note that all these sources are
correctly identified as extended by our algorithm.

We also remove 6 sources since they are easily recognized as blended
by visual inspection.  This effect was included in the simulations
which predicted a number of spurious sources equal to 5.  We argue
that our visual inspection actually remove most of the spurious
detections.  The visual inspection is made possible thanks to the
limited size of our sample.  For larger samples, a more efficient,
self-consistent detection algorithm would be needed in order to keep
the contamination level under control down to lower fluxes, without
recurring to time-consuming visual inspections.  

\begin{figure}
\centering
\includegraphics[width=\columnwidth]{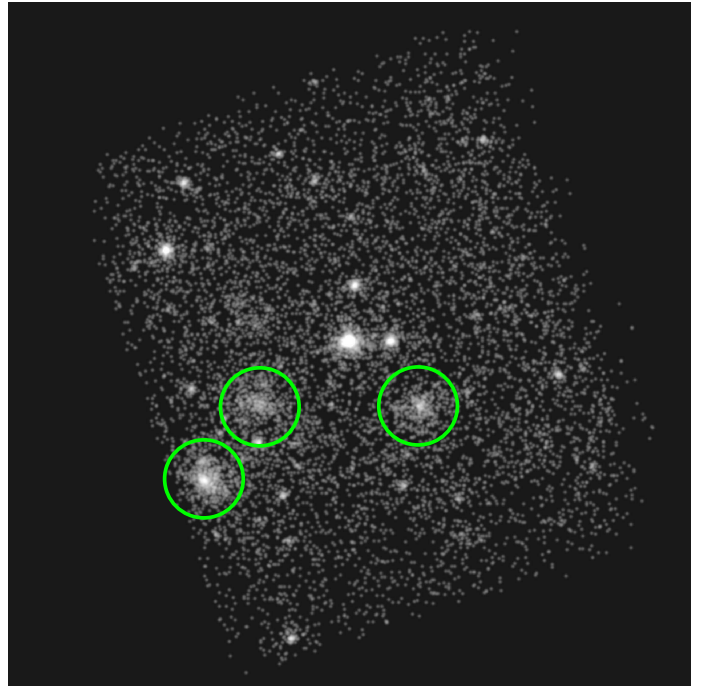}
\caption{XRT smoothed image of the GRB050505 field in the soft
  ($0.5-2.0$ keV) band; the exposure time is $\sim$170 ks, and the image
  size is $\sim$24$\times$24 arcmin. The three clusters in the fields
  are highlighted by the green cicles.  The bright source at the
  center is the GRB.}
\label{xrt_ima}
\end{figure}

Another 3 sources are found to be very close to the edge of the XRT
images, where the gradient of the exposure map significantly affects
the photometry.  We remove them since their photometry would
necessarily rely on a substantial extrapolation of their properties.
The final sample includes 72 extended sources.  Figure~\ref{xrt_ima}
shows a typical example of a GRB field imaged by XRT.  In this case
(GRB050505) three extended sources, highlighted by green circles, have
been selected by our algorithm.

\section{Extended sources photometry} \label{ext_phot}

The measured net counts from each source are computed by aperture
photometry up to an extraction radius R$_{ext}$, after removing the
emission associated to point sources included in this region.
R$_{ext}$ is defined as the radius where the surface brightness
profile of the source reaches the background level. The total net
counts are finally obtained by accounting for the missed flux beyond
$R_{ext}$.  In order to measure the lost signal, we fit the surface
brightness {\it SB} of every extended source with a King profile
modeled as {\it SB}$\propto (1 + (r/r_c)^2)^{3\beta - 1/2}$
\citep{1978Cavaliere} leaving both $r_c$ and $\beta$ free to vary.
Then, we extend the surface brightness profile up to $2 \times
R_{ext}$.  The typical ratio of the total estimated flux to that
actually measured within $R_{ext}$ is $c_f = 1.05-1.10$.
Extrapolating the profile to larger radii has a modest effect on the
final results.  We also checked that the errors on the $r_c$ and
$\beta$ parameters affect the correction factor $c_f$ only at a 10\%
level, and therefore constitute a negligible uncertainty for the final
source flux.

The net count rate measured for each source is obtained dividing the
net counts by the exposure time, after correcting for vignetting
effects in the soft ($0.5-2$ keV) band. The average vignetting
correction in the extraction region is estimated by the ratio of the
photon-weighted average value of the soft exposure map within
R$_{ext}$ to the value of the soft exposure map at the aimpoint.  From
the corrected net count rate, the energy flux is computed simply by
multiplying it by the energy conversion factor (ECF) computed at the
aimpoint for a suitable spectral model.

\begin{figure}
\centering
\includegraphics[width=\columnwidth]{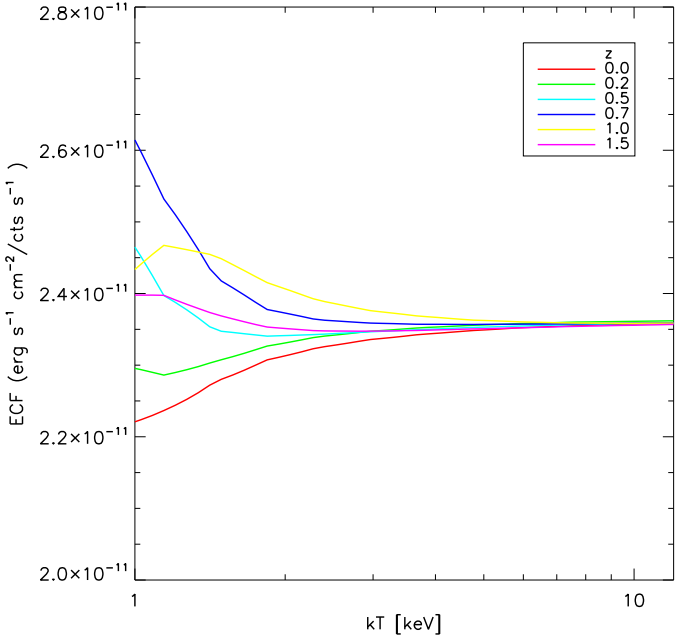}
\caption{Energy Conversion Factor (ECF) as a function of the ICM
  temperature for different redshifts. Here we assumed no Galactic
  absorption.}
\label{ecf_kT}
\end{figure}

The ECF for thermal X--ray emission from the ICM is computed assuming
a {\tt mekal} model within \xspec.  Since we do not know a priori the
temperature of our sources, we explore a range of temperatures from 1
to 12 keV, and a redshift range from 0 to 1.5.  The corresponding ECF
values in the soft band are shown in Figure \ref{ecf_kT} for a typical
Fe abundance of 0.3~$Z_\odot$ in units of \citet{1989Ge}.  We note
that the ECFs vary less than 2\% for \kT$>$3 keV at any redshift. The
largest changes are found for lower temperatures, up to a maximum of
8\% for \kT$=$1 keV. However, given the average $L-T$ relation measured
locally, and considering that it is approximately constant with
redshift
\citep{2011Reichert,2007Branchesi,2006Maughan,2004Ettori,1997Mushotzky},
we can derive a minimum temperature detectable at a given redshift for
our flux limit. If we set $F_{lim} \sim 10^{-14}$ erg s$^{-1}$
cm$^{-2}$, we find that clusters and groups with \kT$<$2 keV can be
observed only for $z<$0.5. This allows us to conclude that the maximum
variation is below 4\%, and that this number is little affected by the
actual ICM metallicity. Therefore we assume an average ECF of $ 2.35
\times 10^{-11}$ erg s$^{-1}$ cm$^{-2}$/ (cts s$^{-1}$) with a maximum
systematic uncertainty of $0.1 \times 10^{-11}$ erg s$^{-1}$
cm$^{-2}$/ (cts s$^{-1}$).

\begin{figure}
\centering
\includegraphics[width=1\columnwidth]{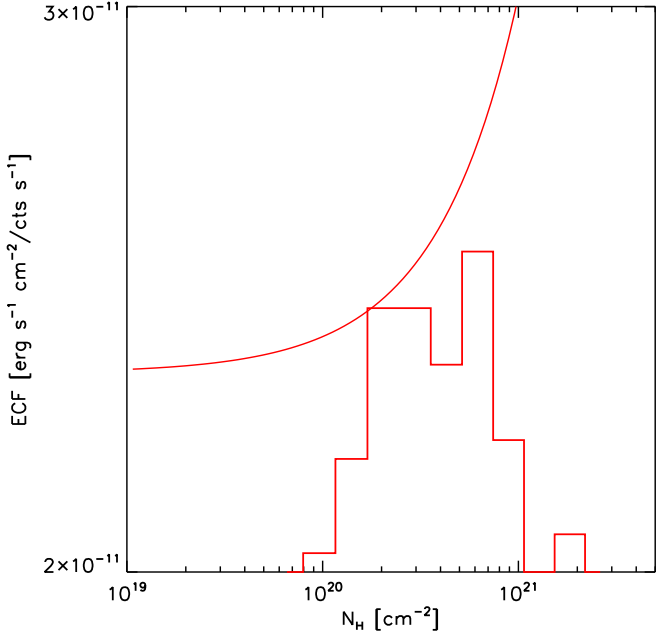}
\caption{Energy Conversion Factor (ECF) as a function of the Galactic
  N$_H$ for a typical cluster (\kT$=$5 keV, $z=$0.4).  The histogram
  shows the distribution of Galactic N$_H$ in our cluster sample.}
\label{ecf_nh}
\end{figure}

The ECFs, however, have a more significant dependence on the Galactic
absorption.  This can be estimated thanks to the Galactic N$_H$ values
measured in the radio survey in the Leiden/Argentine/Bonn survey
\citep{2005LAB}. As shown by the histogram in Figure \ref{ecf_nh}, the
distribution of Galactic N$_H$ for our sources implies changes up to
the order of 30\%.  Therefore we use for each detected source the ECF
appropriate to the corresponding field.  We note that self shielding
of molecular Hydrogen from ambient UV radiation is expected to occur
for $N_H> 5\times 10^{20}$ cm$^{-2}$ \citep{1999arab}, and this
molecular gas absorbs also X--rays.  Therefore for this $N_H$ values X--ray
fluxes are slightly underestimated.  We do not correct for this effect
\citep[see also ][]{2011lloyd-davies}.

To summarize, the total unabsorbed, soft fluxes, including the
correction up to 2~$R_{ext}$, are measured as:

\begin{equation}
S_{0.5-2 keV} = ECF(N_H) \times c_f \times C_{rate}
\end{equation}

\noindent
where $ECF(N_H)$ is the energy conversion factor which accounts for
the Galactic absorption (see Figure~\ref{ecf_nh}), $c_f$ is the
correction factor for the flux between $2\times R_{ext}$ and
$R_{ext}$, and $C_{rate}$ is the soft band count rate mesured within
$R_{ext}$ and corrected for vignetting effects.

\section{Cluster catalog and Number Counts}\label{cluster_cat}

Aperture photometry within $R_{ext}$ and total energy flux are shown
in Table \ref{table2} for all the sources of our sample.  The net
counts error is obtained from the Poissonian error on the numbers of
net photons.  We remark that, as explained in detail in the previous
section, the photometry of our extended sources refers to the
extraction radius R$_{ext}$ defined as the radius where the fitted
surface brightness falls below the measured local background, while
the total energy flux include the correction factor $c_f$ for the flux
lost beyond R$_{ext}$. 

\begin{figure}
 \centering
 \includegraphics[width=\columnwidth]{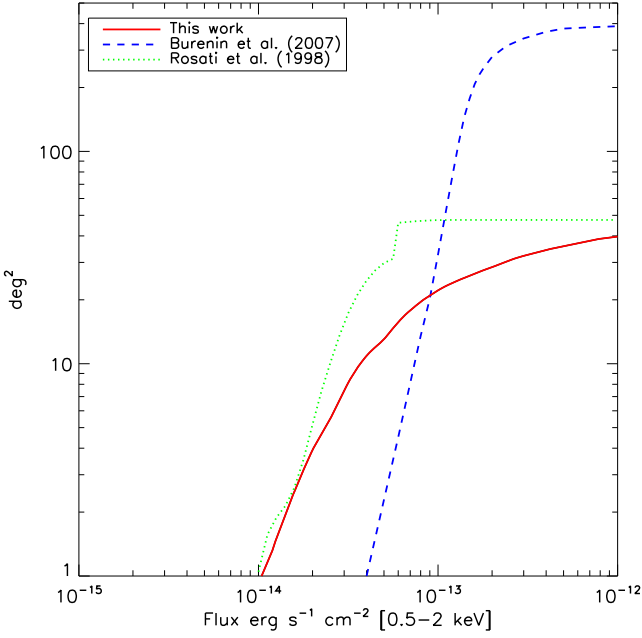}
 \caption{Solid angle $\Omega$ covered by our survey as a function
   of the soft flux. As a comparison, we show also the sky coverage of
   the 400 Square Degree Survey \protect\citep[blue dashed
   line]{2007Burenin} and the sky coverage of the RDCS \protect\citep[green
   dotted line]{1998Rosati}.}
 \label{skycov}
 \end{figure}

In order to compute the number counts for our cluster sample, we need
to derive the sky coverage of our survey.  The sky coverage is
determined by the sharp limit in the total net photons.  For each
field we compute a flux-limit map obtained as $ECF(N_H)\times
100/Expmap(t)$, where $Expmap(t)$ is the exposure map in units of
effective time, therefore including the effect of vignetting.  Then,
the solid angle covered by the survey above a given flux is obtained
by measuring the total solid angle where the flux-limit is lower than
a given flux.  In Figure \ref{skycov} we show the resulting sky
coverage $\Omega (S)$ and compare it with the sky coverage of the 400
Square Degree ROSAT PSPC Galaxy Cluster Survey by \citet{2007Burenin}
and with the sky coverage of the ROSAT Deep Cluster Survey (RDCS) by
\citet{1998Rosati}. Unfortunately, sky coverage curves for other
on-going XMM/\chandra\ surveys have not been published.  The cumulative
number counts for sources brighter than a given flux S is therefore
given by

\begin{equation}
N(>S) = \Sigma_{c_f\times S_i>S} C_i^{-1}/\Omega(S_i)
\label{lnls}
\end{equation}

\noindent
where $S$ is the total flux, $S_i$ is the soft flux within $R_{ext}$
of the $i^{th}$ source, $c_f$ is the correction factor for the flux
outside $R_{ext}$ and $\Omega(S_i)$ is the solid angle corresponding
to $S_i$.  Finally, $C_i$ is the completeness factor plotted in Figure
\ref{completeness} which depends on the net detected photons of the
$i^{th}$ source.  In this way, each source is weighted with a factor
inversely proportional to the survey completeness which depends only
on source photometry.

An alternative procedure to correct for completeness is obtained by
randomly extracting the missing sources and computing their average
effect.  The procedure consists in dividing the sources in 3 bins of
net photons (100-150, 150-200, 200-300).  Then, in each bin we
randomly add a number of sources drawn from a Poissonian distribution
centered on the expected number of missed sources according to Figure
\ref{completeness}.  Finally we assign a random exposure and energy
conversion factor among those in the survey to each mock source, and
compute its energy flux.  Finally the logN-logS is recomputed for each
Monte Carlo realization.  The average logN-logS obtained with this
procedure with $10^4$ realizations is in very good agreement with that
obtained with equation \ref{lnls}, therefore confirming that our
treatment of the completeness is robust.

\begin{figure}
\centering
\includegraphics[width=\columnwidth]{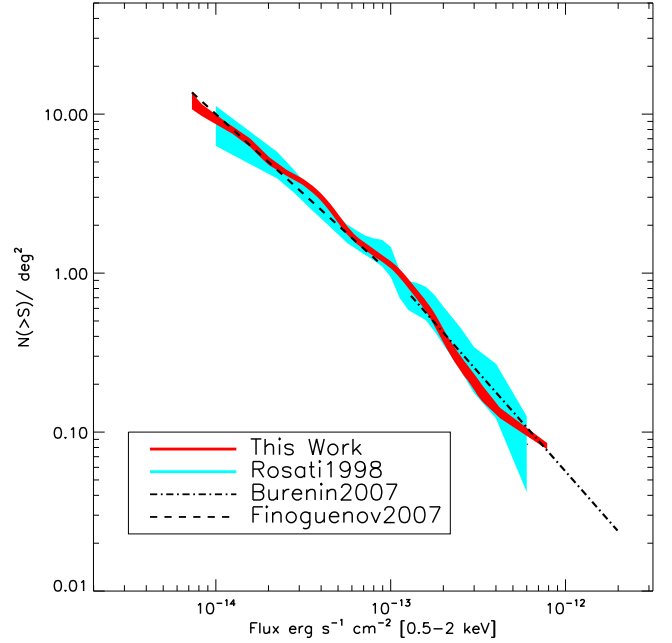}
\caption{Cumulative number counts, corrected for completeness, with 1
  $\sigma$ confidence level for SXCS (red area), compared to the
  logN--logS derived from the ROSAT Deep Cluster Survey (cyan area),
  \protect\citet{1998Rosati}.  Dot-dashed and dashed lines show the
  fit to the 400d \protect\citep{2007Burenin} and to the COSMOS
  \protect\citep{2007Finoguenov} number counts, respectively.}
\label{lognlogs}
\end{figure}

Our logN-logS is shown in Figure \ref{lognlogs} with its 68\%
confidence limits (shaded red region). The confidence limits are
computed with Monte Carlo realizations. We re-extract the flux of each
source  times assuming its Poissonian uncertainties on the net
detected counts, including the systematic uncertanty on the ECFs,
obtaining $10^{4}$ realizations of the logN-logS. Then, at each flux, we
compute the 68\% confidence interval around the mean value.

Our results are consistent with the logN-logS measured with the ROSAT
Deep Cluster Survey \citep[RDCS,][]{1998Rosati}, shown in Figure
\ref{lognlogs} as the cyan region, down to fluxes $\sim 10^{-14}$ erg
s$^{-1}$ cm$^{-2}$.  We are also consistent with the the logN-logS
measured in the 400 deg$^2$ survey \citep{2007Burenin} and in the
COSMOS field \citep{2007Finoguenov}, shown as the dash-dotted and
dotted lines respectively.

\begin{figure}
\centering
\includegraphics[width=\columnwidth]{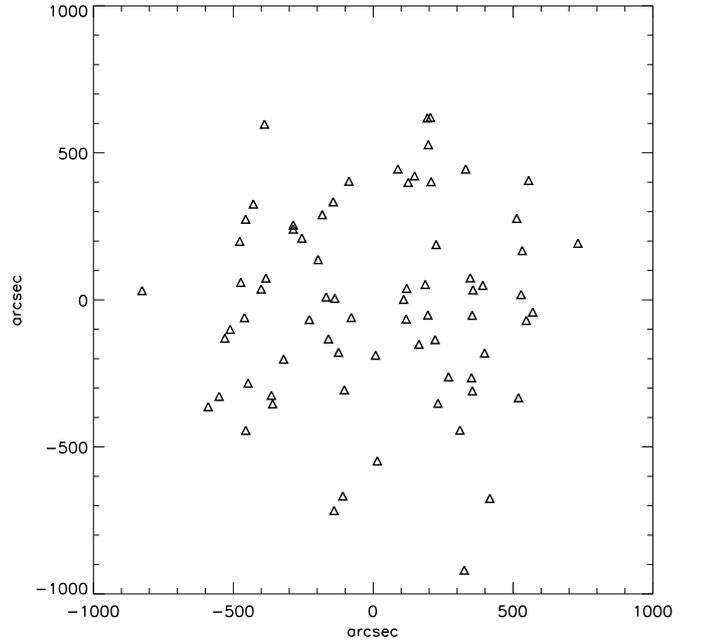}
\caption{Positions of our sources with respect to the GRB position
  (set to 0,0). }
\label{clustering}
\end{figure}

The good agreement with previous results confirm the proper
characterization of our sample.  Nevertheless, we also perform a test
against a possible correlation between the GRB position and the
position of our sources, since the detection of an enhanced density of
clusters and groups toward the position of GRB would bias our survey.
In Figure \ref{clustering} we plot the positions of the sources in our
sample with respect to the GRB positions (set to 0,0).  We do not find
any hint of an increasing surface density of extended sources towards
the GRB position.  We also tested for such a possible bias by
measuring the logN-logS for the sources whose distance is,
respectively, below 400 arcseconds from the GRB and between 400 and
800 arcseconds from the GRB.  The two logN-logS agree with each other
within the uncertainties, showing that there is no correlation between
GRB and clusters \citep[see also][]{2007Berger}.  We can safely assume
that our field selection is unbiased with respect to X--ray clusters.

In order to check whether we properly treated the effect of the
Galactic absorption, we split the survey into two segments according
to the Galactic absorption, above and below $N_H= 3 \times 10 ^{20}$
cm$^{-2}$.  We find that the logN-logS computed in the two cases are
in very good agreement, and therefore we conclude that our flux
measurements do not appear to suffer any bias from Galactic
absorption.

To summarize, the sources included in the SXCS catalog represent a
sample of group and cluster candidates with a negligible
contamination, a well defined selection function and a robustly
estimated completeness, spanning two orders of magnitude in flux, and
reaching the flux limit of $\sim 10^{-14}$ erg s$^{-1}$ cm$^{-2}$.

\section{Cross-correlation with X--ray and optical catalogs.}\label{cross_corr}

We checked for counterparts both in previous X--ray surveys and in
optical cluster surveys, assuming a search radius of 1 arcmin from our
X--ray centroid.  The results are shown in Table \ref{counterpart}.
We find 9 X--ray counterparts to our sources in ROSAT, ASCA and
\chandra\ catalogs, none of them characterized as extended and none of
them with a published redshift.  We also find a total of 20 previously
known, optically identified clusters. Among them, 6 are found in the
Wen+Han+Liu cluster sample \citep[WHL,][]{2009Wen}, which is an
optical catalog of galaxy clusters obtained from an adaptive--matched
filter finder applied to Sloan Digital Sky Survey DR6.  We find 4
clusters in the \citet{2011Szabo} catalog (AMF) based on a similar
method in the SDSS DR6.

We also report one cluster from the Gaussian Mixture Brightest Cluster
Galaxy \citep[GMBCG,][]{2010Hao} based on SDSS DR7, which is an
extension of the maxBCG cluster catalog \citep{MaxBCG} to redshift
beyond 0.3 and on a slightly larger sky area ($\sim 8000$ deg$^2$),
and one cluster in the MaxBCG catalog itself.  

Finally, we find 3 clusters in the Abell catalog \citep{1989Abell}, two
cluster in the Northern Sky optical Cluster Survey \citep[NSCS,][]{2003NSC,
  2004NSC}, and one cluster in each of the following catalogs: the
Zwicky Cluster Catalog \citep[CGCG,][]{1963Zwicky}, the Sloan Digital
Sky Survey C4 Cluster Catalog \citep[SDSS-C4-DR3, based on
  DR3,][]{2005SDSS-C4-DR3, 2007SDSS-C4-DR3}, the Edinburgh-Durham
Southern Galaxy Catalogue \citep[EDCC,][]{1992EDCC}.  Three out of
twenty clusters also have an X--ray counterpart.

To increase the number of available redshifts, we also search for
galaxies with published redshift not associated to previously known
clusters, within a search radius of 7 arcsec from the X--ray centroid
of our sources.  In Table \ref{counterpart} we report 11 galaxies with
redshift, within a search radius of 7 arcsec, as a complement to the
redshift of the cluster counterpart. The redshift of the central
galaxy candidate is always consistent with the photometric redshift of
the optical cluster counterpart when present.  In the 4 cases where no
optical cluster counterpart is found, we tentatively assign the galaxy
redshift to our X--ray source.  We note that both cluster and central
galaxy counterparts have been found by adopting a simple matching
criterion.  A more refined analysis of the optical data covering our
survey is under way (Tundo et al. in preparation).

\begin{figure*}[h]
 \centering
\includegraphics[width=0.5\columnwidth]{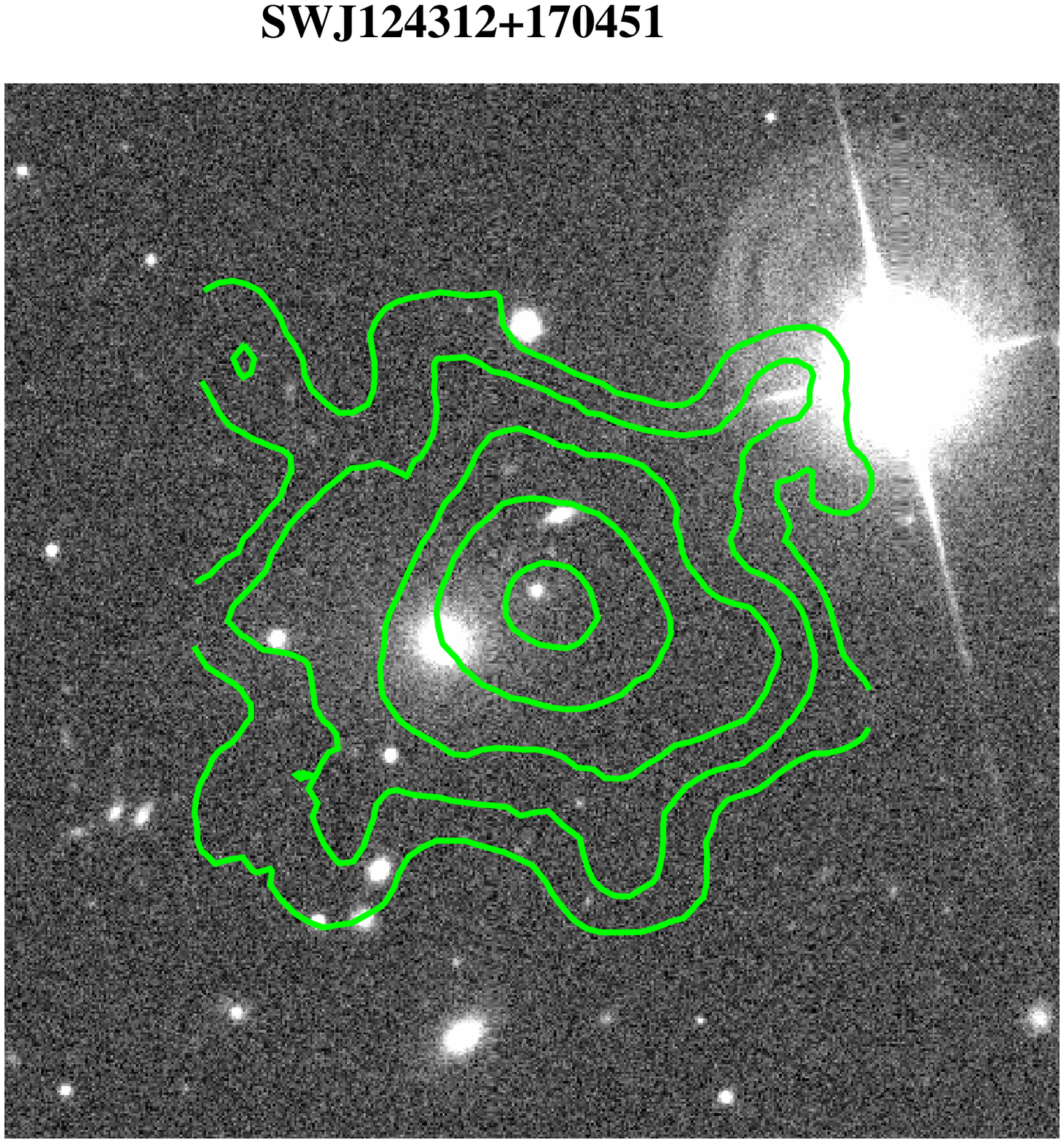}
\includegraphics[width=0.5\columnwidth]{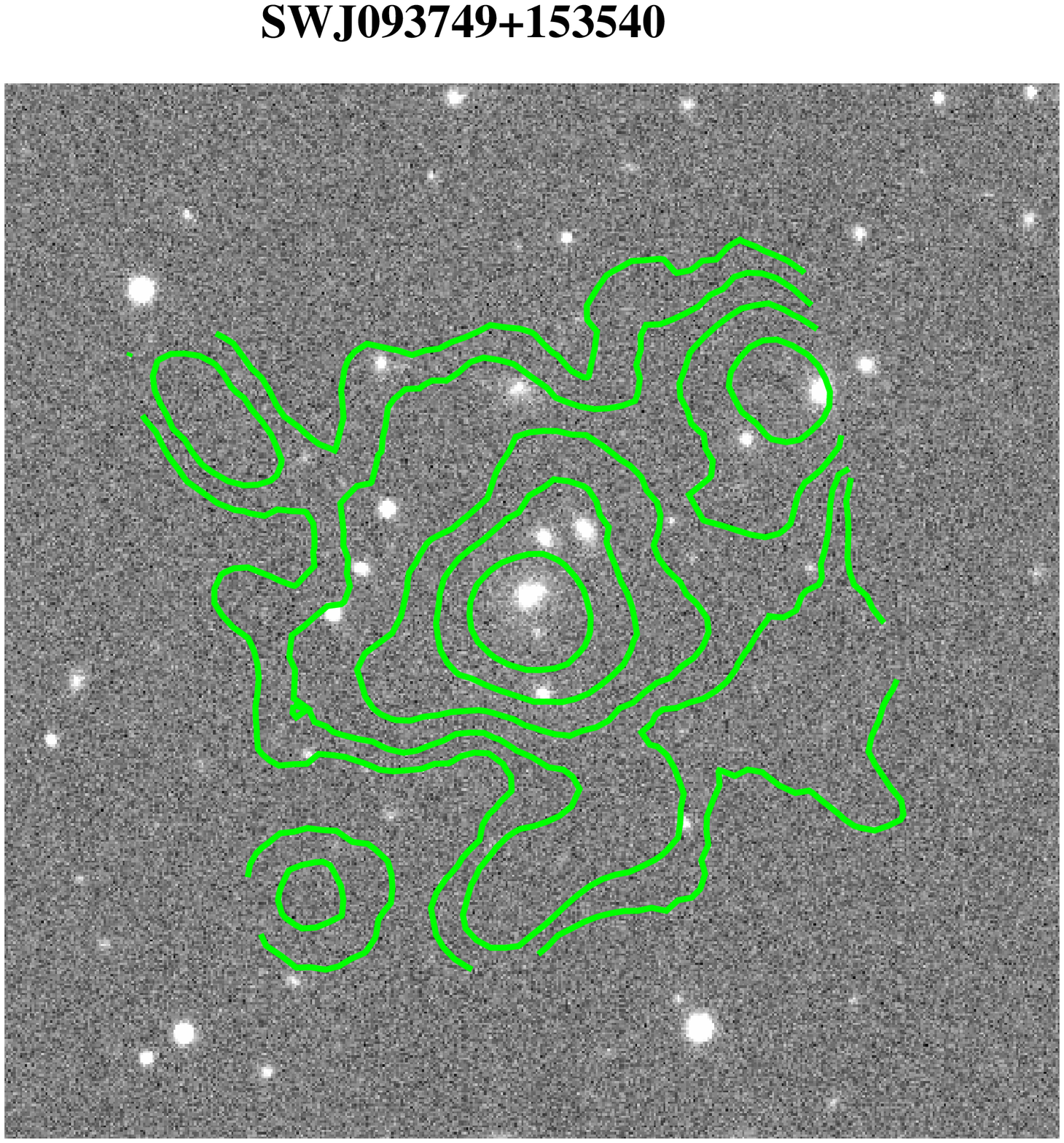}
\includegraphics[width=0.5\columnwidth]{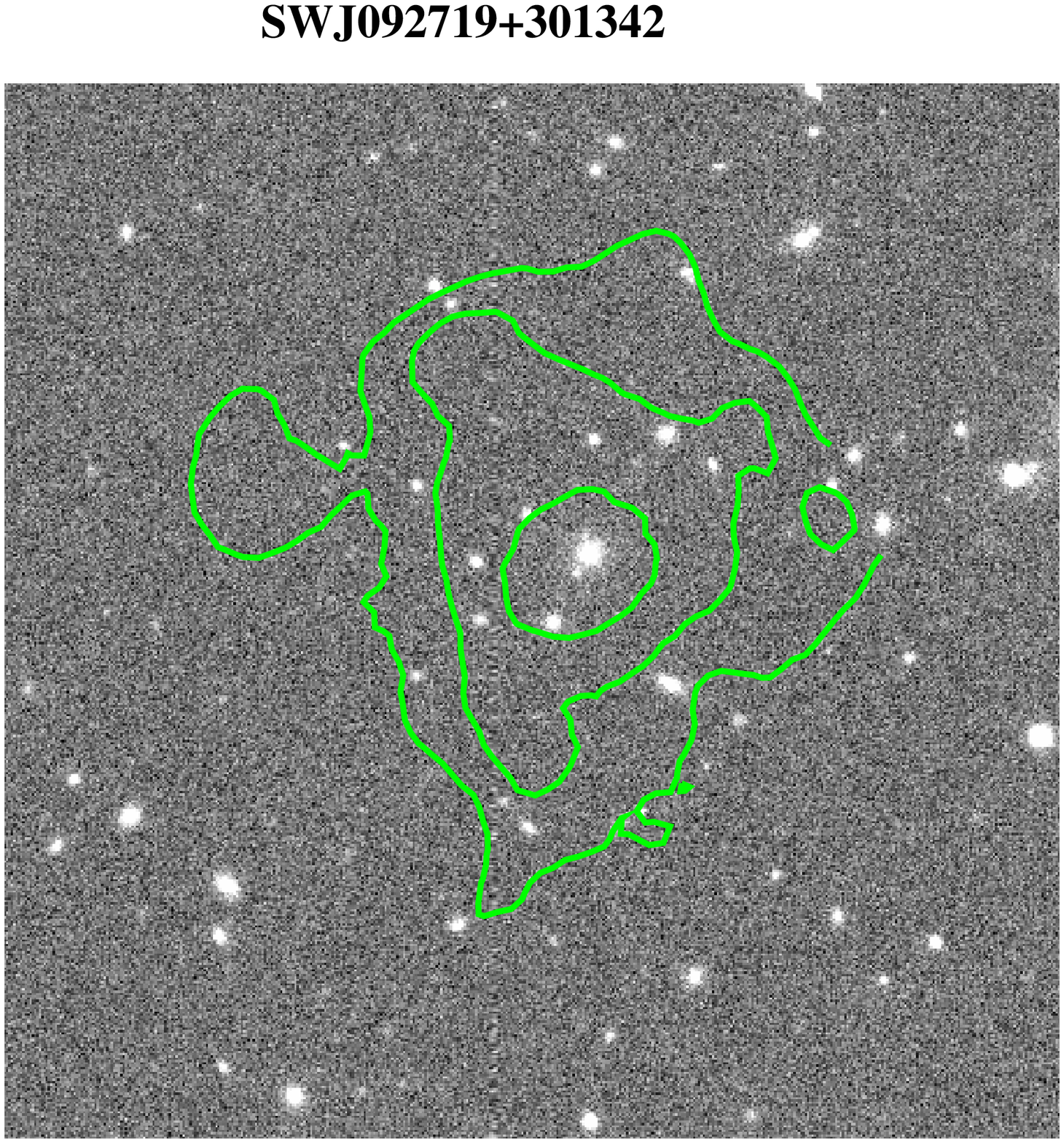}
\includegraphics[width=0.5\columnwidth]{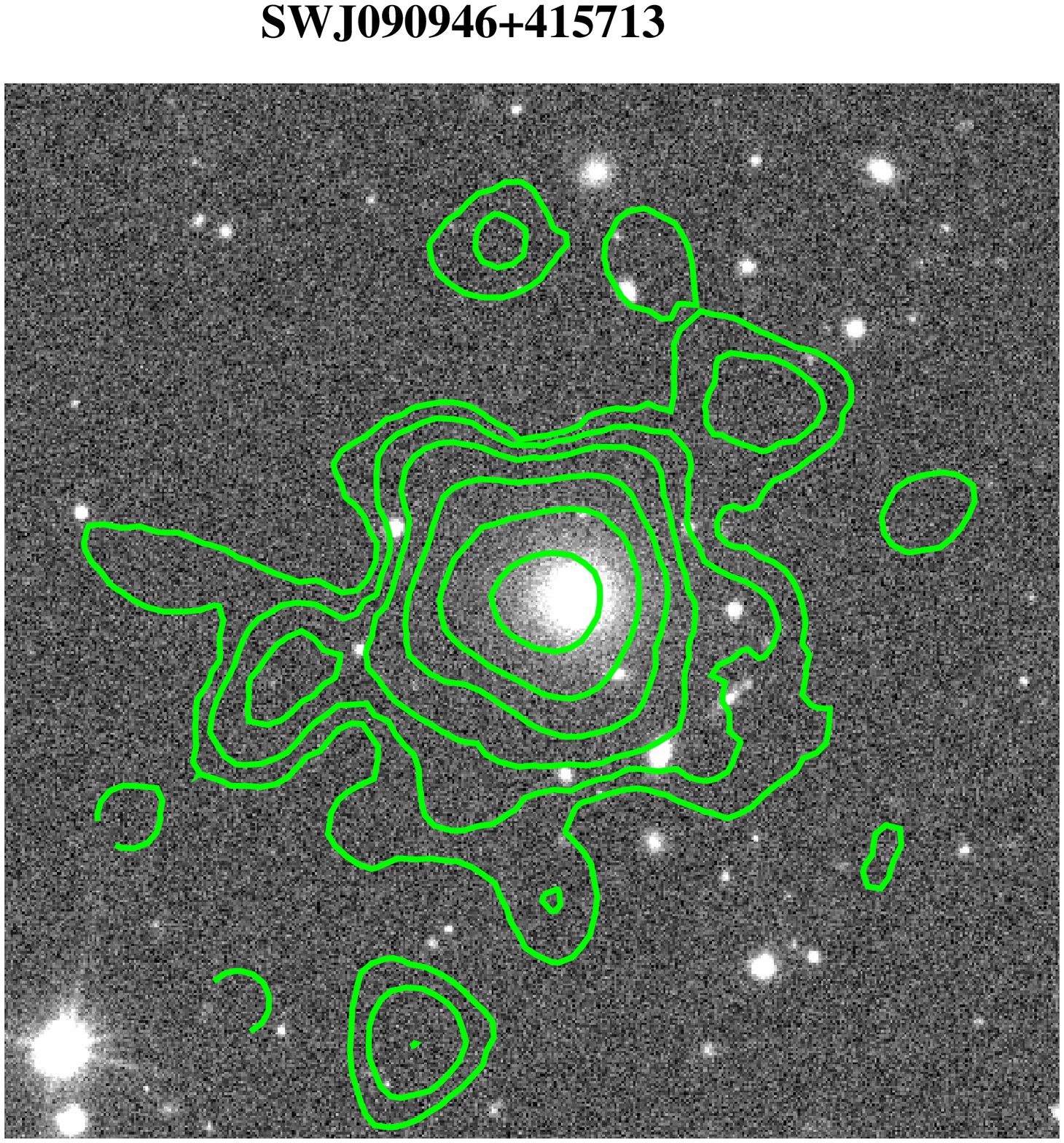}
\includegraphics[width=0.5\columnwidth]{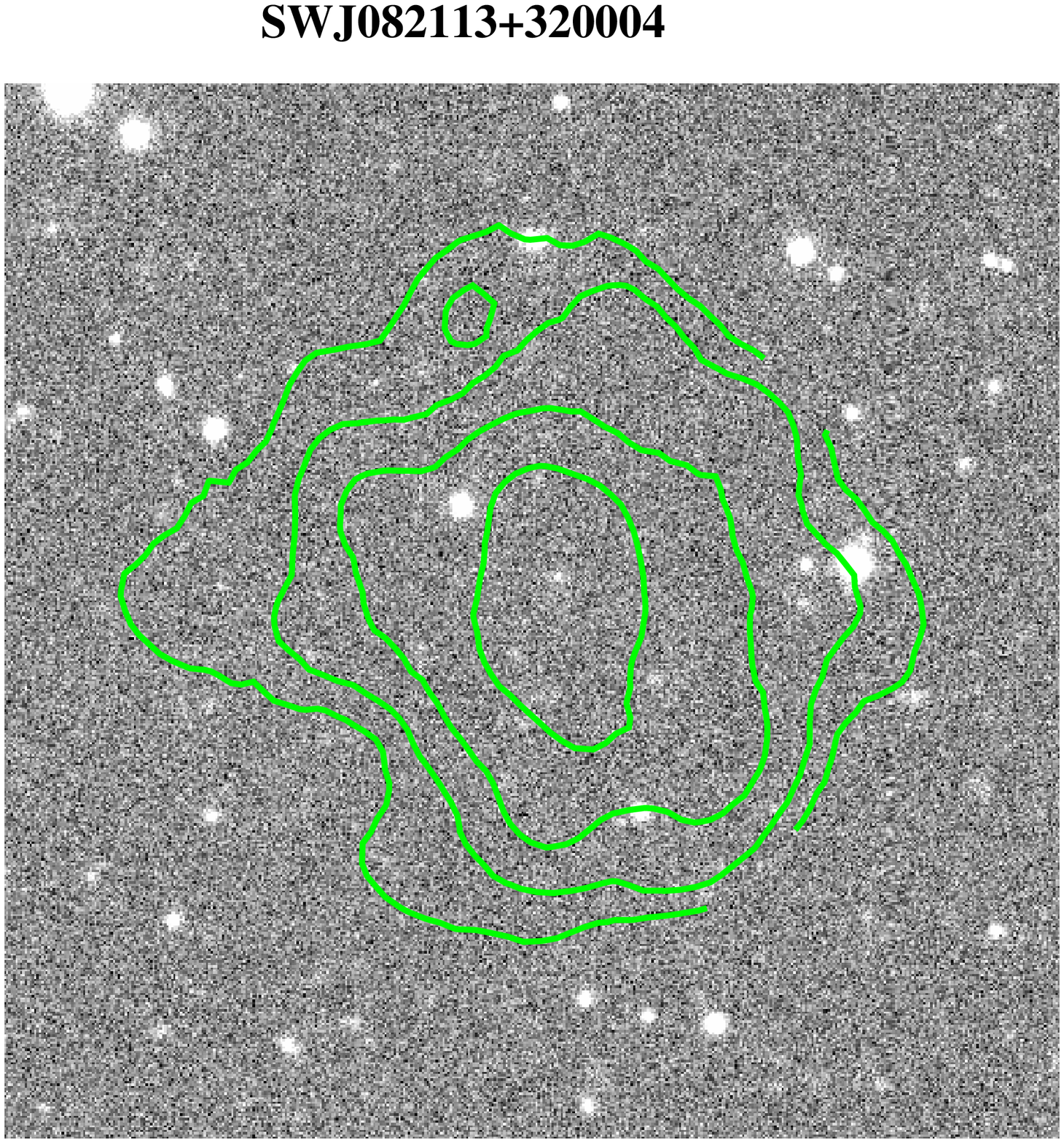}
\includegraphics[width=0.5\columnwidth]{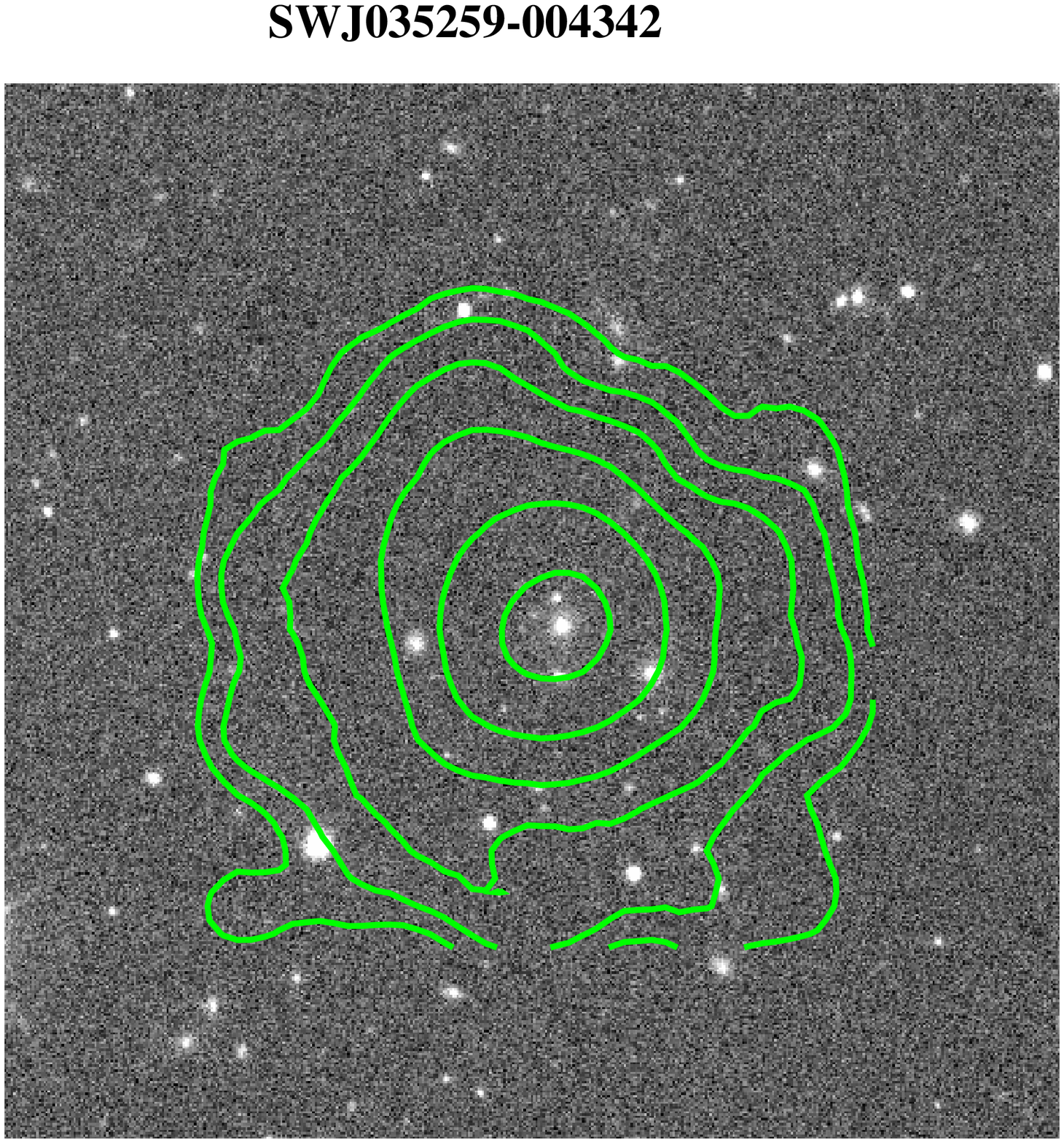}
\includegraphics[width=0.5\columnwidth]{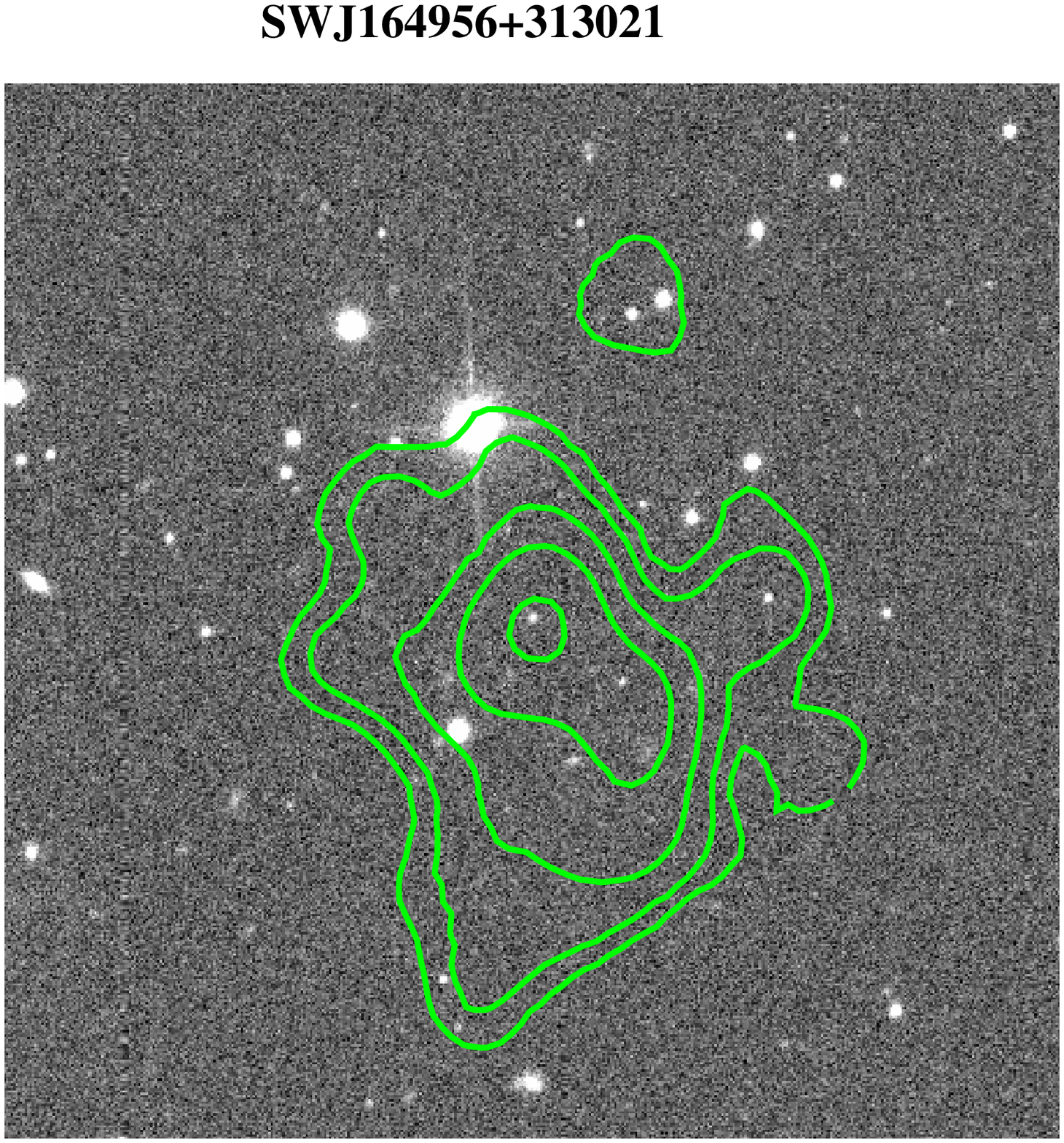}
\includegraphics[width=0.5\columnwidth]{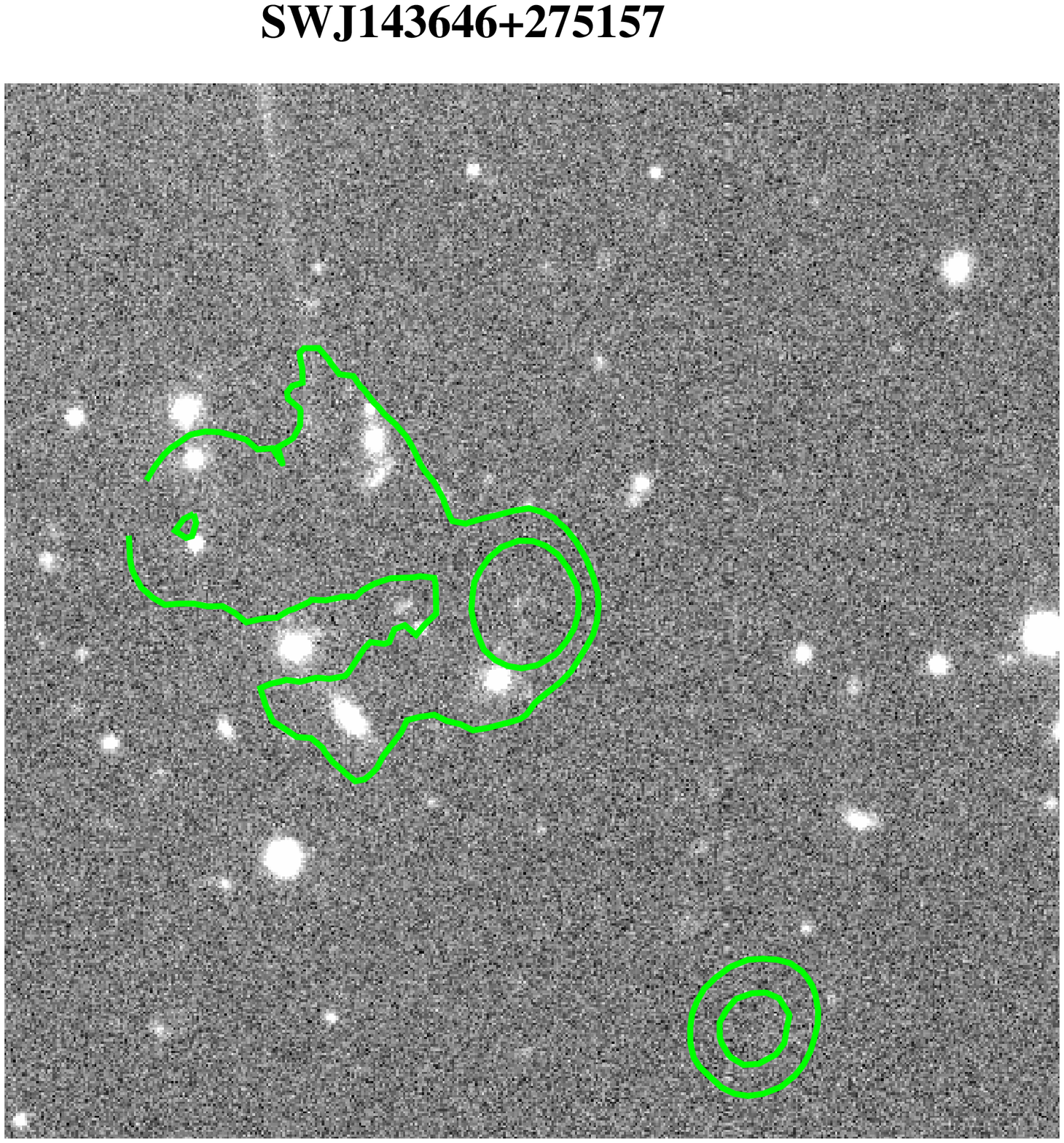}
\caption{Selection of SXCS cluster candidates imaged by the Sloan
  Digital Sky Survey (SDSS) with X--ray contours overlaid in
  green. The image size is 3$\times$3 arcmin. The X--ray contours are
  computed on XRT images in the $0.5-2$ keV energy band with the
  \chandra\ Ciao software, and the contours are drawn at respectively
  9, 15, 30, 50, 90, 160 and 300 $\sigma$, where $\sigma$ is given by
  the local background level.}
 \label{X_contours}
 \end{figure*}

To summarize, we have 24 optical redshifts (spectroscopic or
photometric) published in the literature and associated to our cluster
candidates.  Overall, 46 sources in our catalog are new detections,
both as X--ray sources and as clusters of galaxies.  Finally, we remark
that a total of 32 SXCS sources fall in  SDSS fields.  Thanks to
the SDSS depth, we expect to increase the identification of our
cluster candidates.  In particular we expect to be able to identify at
least the brightest cluster galaxy up to $z\leq 1$ for those cluster
candidates which are not already included in the SDSS catalogs.  This
will also provide an estimate of the photometric redshift of the host
cluster.  Figure \ref{X_contours} shows a selection of SDSS r-band
images of SXCS fields, with X--ray contours overlaid in green.

We stress that we can also attempt a measure of the source redshift
through the X--ray spectral analysis for a significant fraction of the
sample.  The requirements for a successfull identification of the
redshifted $K_\alpha$ Fe line, shown in \citet{2011Yu} for \chandra,
do not apply to most of the SXCS sources.  However the lower
background and the flatter effective area of XRT may
allow X--ray redshift measurements in a lower S/N regime.  A
preliminary result indicates that the X--ray redshift $z_X$ of about
30\% of the SXCS sources can be successfully measured.  This is an
important aspect since the measure of $z_X$ will complement the
photometric and spectroscopic redshifts obtained with optical
follow--up with the aim of using the cluster sample for cosmological
tests.  We will present the spectral analysis in a companion paper
(Moretti et al. in preparation).

\section{Future prospects}\label{discus}

SXCS data have different properties from \chandra\ and XMM--Newton
data as described in \S~\ref{sur_sec} and \S~\ref{field_sel}, and
therefore SXCS extended sources have a different selection function.
For these reasons the SXCS is a valuable touchstone to the cluster
catalogs obtained from XMM-Newton and \chandra.  The full SXCS survey
(Liu et al. in preparation) is expected to cover a larger solid angle
especially at bright fluxes, and to be at least a factor of two
deeper.  The exploitation of the \chandra\ and XMM archives is still
far from being concluded, and we expect to see the number of high--$z$
clusters ($z>1$) to double in the next years, and in particular to see
the increase of the sample of clusters at $z>1.5$ which are being
found only recently.  However, a major breakthrough in the field of
cluster surveys can be achieved only with a survey dedicated mission.

The planned eROSITA satellite \citep{2010Predehl,2010Cappelluti} and
the proposed WFXT mission \citep{2010WFXT2, 2010Murray} can provide a
large number of new detections and X--ray spectra for a large number
of them.  The eROSITA mission is expected fo fly in the near future,
and it will finally provide an all-sky survey almost two orders of
magnitude deeper than the last one performed by the ROSAT satellite
\citep{1999voges}, filling a gap of almost 20 years.  Unfortunately,
eROSITA will be confusion limited in the flux regime where the
majority of the high--$z$ clusters are expected.  In addition, the low
hard--band sensitivity (the effective area decreases an order of
magnitude\footnote{http://www.mpe.mpg.de/erosita/} from 1 keV to 4
keV) hampers a measure of the temperature for hot clusters, limiting
cosmological studies.

The WFXT is the only X--ray mission optimized for efficient wide-area
surveys, with a design which yields good angular resolution and a
constant image quality across a large (1 deg$^2$) field of view
\citep[see][]{2010WFXT2}. These aspects, coupled with a large
effective area, will provide not only a survey more than one order of
magnitude deeper than eROSITA, but a direct measurement of global
temperatures, density profiles and redshifts for a significant
fraction of the cluster sample, thus allowing cosmological studies
independently of optical and spectroscopic follow--up
\citep{2010Borgani}. Simulations based on the original WFXT design
\citep{2010WFXTsimu} show that such a mission is able to match in
depth, survey volume and angular resolution surveys at other
wavelengths which are planned for the next decade.

\section{Conclusions}\label{conc}

We present a new sample of X--ray selected groups and clusters of
galaxies obtained with the X--ray Telescope (XRT) on board of the
Swift satellite.  We search for extended sources among 336 GRB fields
imaged with XRT with galactic latitude $|b|>$20$^\circ$ (available in
the archive as of April 2010). We identify extended sources with a
simple criterion based on the measurement of the HPR within a box of
$45\times 45$ arcsec of the source image.  We apply a sharp threshold
of 100 net soft counts within the extraction radius $R_{ext}$ to
select reliable extended sources.  Extensive simulations showed that
our method, despite being very simple, provide us with an X--ray sample
with an expected high completeness and a low contamination, also
thanks to a careful {\sl a posteriori} visual inspection.

Our final group and cluster catalog consists of 72 X--ray sources.
The sky coverage of the survey goes from the total 40 deg$^2$ to 1
deg$^2$ at a flux limit of about $10^{-14}$ erg s$^{-1}$ cm$^{-2}$.
The corresponding logN--logS is in very good agreement with previous
deep surveys.

We directly verified that there is no correlation between the position
of the clusters and that of the GRBs. In other words, selecting our
cluster sample from XRT GRB fields does not result in any spatial bias
with respect to the GRB positions.

A cross correlation with X--ray catalogs shows that only 9 SXCS
sources were previously identified in the X--ray band, none of them
classified as extended.  A search in optical databases (mostly based
on SDSS data) allows us to find the counterparts of 20 clusters.  In
addition, 4 galaxies with redshift have been found to be within 7
arcsec from the X--ray centroid, and therefore they are considered as
possible identification of the central galaxy of the group/cluster
candidate.  Overall, only 20 sources are confirmed as clusters using
data from the literature, while 30 sources have some counterpart in
the NASA Extragalactic Database, and, finally, 42 sources are newly
identified.

We estimate that about one third of the sample is detected with a S/N
high enough to allow the measure of the redshift from the X--ray
spectral analysis \citep[see][]{2011Yu}, as we will show in a
companion paper (Moretti et al. in preparation).  Thanks to the
quality of our X--ray data, and thanks to the synergy with other
surveys, specifically the SDSS, we are able to provide a X--ray cluster
sample with a well established completeness function down to a flux
limit comparable to that of the deepest cluster surveys based on ROSAT
data \citep[RDCS, see][]{1998Rosati}, and with a comparable
statistics.  In addition, we expect to detect a few clusters with
redshift $z\geq$1.  Overall, the SXCS is expected to give a
significant contribution in the field of X--ray clusters surveys also
thanks to its peculiar properties of a low background and a constant
PSF.  These properties are two key requirements for  future
wide--area, X--ray surveys, as foreseen by proposed future 
missions aiming at bringing the X--ray sky to the same depth and
richness of the optical and IR sky in the next decade.
A deeper, extended release of the SXCS based on a new detection
algorithm tailored to XRT images is currently undergoing (Liu et
al. in preparation).  Catalog and data products of SXCS, constantly
updated, are made avalilable to the public through the website {\tt
  http://adlibitum.oats.inaf.it/sxcs}.

\acknowledgements We acknowledge support from ASI-INAF I/088/06/0 and
ASI-INAF I/009/10/0.  PT aknowledges support under the grant INFN
PD51. GT and SC acknowledge support from ASI-INAF I/011/07/0.  We
thank the anonymous referee for helping us to improve sgnificantly our
paper with extremely detailed and useful comments.  This research has
made use of the NASA/IPAC Extragalactic Database (NED) which is
operated by the Jet Propulsion Laboratory, California Institute of
Technology, under contract with the National Aeronautics and Space
Administration.

\bibliography{references_Tundo2012}

\longtabL{2}{
\begin{landscape}
\begin{longtable}{lcccccccc}
\caption{\label{table2} SXCS sources.}\\
\hline\hline
\bf Name & \bf ra & \bf dec             & \bf Exptime & \bf  N$_H$            & \bf R$_{ext}$ & \bf Soft Cts & \bf Soft SNR & \bf Soft Flux          \\ 
\bf      & \multicolumn{2}{c} {\bf deg} & \bf [s]     & \bf $10^{20}$ cm$^{-2}$ & \bf arcsec   & \bf          & \bf          & {\bf 10$^{-14}$erg s$^{-1}$ cm$^2$} \\ 
\hline
\endfirsthead
\caption{continued.}\\
\hline\hline
\bf Name & \bf ra & \bf dec             & \bf Exptime & \bf  N$_H$            & \bf R$_{ext}$ & \bf Soft Cts & \bf Soft SNR & \bf Soft Flux          \\ 
\bf      & \multicolumn{2}{c} {\bf deg} & \bf [s]     & \bf $10^{20}$ cm$^{-2}$ & \bf arcsec   & \bf          & \bf          & {\bf 10$^{-14}$erg s$^{-1}$ cm$^2$} \\ 
\hline
\endhead
\hline
\endfoot
SWJ000345-530149 &   0.93793 & -53.03028 &       301809 &  1.60 &  18 &    191$\pm$  17 &  10.8 &   1.6$\pm$  0.2 \\
SWJ000315-525510 &   0.81655 & -52.91970 &       303488 &  1.60 &  44 &   1487$\pm$  46 &  31.9 &  12.0$\pm$  1.3 \\
SWJ000324-525350 &   0.85004 & -52.89743 &       289366 &  1.60 &  52 &   1847$\pm$  52 &  35.2 &  15.7$\pm$  1.9 \\
SWJ002437-580353 &   6.15751 & -58.06475 &        74764 &  1.22 &  43 &    315$\pm$  20 &  15.5 &  10.3$\pm$  1.2 \\
SWJ003316+193922 &   8.31936 &  19.65637 &        44186 &  4.14 &  41 &    126$\pm$  13 &   9.4 &   7.5$\pm$  1.1 \\
SWJ005500-385226 &  13.75198 & -38.87410 &        40951 &  3.32 &  36 &    104$\pm$  12 &   8.3 &   6.6$\pm$  1.0 \\
SWJ011432-482824 &  18.63735 & -48.47342 &        38003 &  1.95 &  47 &    158$\pm$  15 &  10.5 &  10.3$\pm$  1.4 \\
SWJ012210-130422 &  20.54544 & -13.07302 &        62075 &  2.35 &  28 &    101$\pm$  12 &   8.3 &   4.1$\pm$  0.7 \\
SWJ012302+375615 &  20.76171 &  37.93769 &        75637 &  6.03 &  31 &    120$\pm$  12 &   9.3 &   4.4$\pm$  0.7 \\
SWJ015753+165933 &  29.47116 &  16.99270 &        77327 &  4.85 &  27 &    104$\pm$  11 &   8.8 &   3.6$\pm$  0.6 \\
SWJ020744+002055 &  31.93682 &   0.34872 &        83492 &  2.34 &  24 &    119$\pm$  12 &   9.7 &   3.6$\pm$  0.5 \\
SWJ021705-501409 &  34.27108 & -50.23602 &       131772 &  1.80 &  44 &    749$\pm$  30 &  24.4 &  14.0$\pm$  1.6 \\
SWJ021747-500322 &  34.44821 & -50.05631 &       122374 &  1.80 &  29 &    207$\pm$  16 &  12.3 &   4.2$\pm$  0.6 \\
SWJ022344+382311 &  35.93546 &  38.38647 &       150608 &  4.40 &  13 &    102$\pm$  11 &   9.2 &   1.8$\pm$  0.3 \\
SWJ022546-185553 &  36.44202 & -18.93140 &       403267 &  2.81 &  37 &    813$\pm$  33 &  24.3 &   5.1$\pm$  0.6 \\
SWJ023224-712020 &  38.10299 & -71.33891 &       316584 &  5.79 &  19 &    212$\pm$  17 &  11.9 &   1.8$\pm$  0.2 \\
SWJ023302-711634 &  38.26142 & -71.27616 &       323540 &  5.79 &  43 &   1284$\pm$  42 &  30.2 &  10.8$\pm$  1.2 \\
SWJ023924-250504 &  39.85054 & -25.08445 &       118554 &  2.23 &  38 &    323$\pm$  21 &  15.3 &   6.8$\pm$  0.9 \\
SWJ024010-251121 &  40.04202 & -25.18923 &       242137 &  2.23 &  19 &    115$\pm$  13 &   8.5 &   1.2$\pm$  0.2 \\
SWJ035259-004342 &  58.24825 &  -0.72847 &       107623 & 11.30 &  44 &   1461$\pm$  40 &  36.3 &  42.2$\pm$  4.4 \\
SWJ035310+213335 &  58.29573 &  21.55977 &        83256 & 11.30 &  32 &    137$\pm$  14 &   9.8 &   5.1$\pm$  0.8 \\
SWJ044144-111536 &  70.43740 & -11.26022 &        68676 &  4.61 &  36 &    311$\pm$  19 &  15.7 &  12.0$\pm$  1.5 \\
SWJ062155-622834 &  95.48190 & -62.47626 &      1039776 &  4.49 &  13 &    286$\pm$  20 &  13.8 &   0.7$\pm$  0.1 \\
SWJ082113+320004 & 125.30467 &  32.00130 &       117078 &  3.35 &  35 &    597$\pm$  27 &  22.0 &  13.1$\pm$  1.4 \\
SWJ083340+331102 & 128.41815 &  33.18397 &        32951 &  4.07 &  39 &    142$\pm$  13 &  10.7 &  11.3$\pm$  1.6 \\
SWJ084749+133141 & 131.95442 &  13.52810 &        77386 &  3.18 &  54 &   3824$\pm$  63 &  60.6 & 126.4$\pm$ 12.9 \\
SWJ085524+110201 & 133.85220 &  11.03384 &        85418 &  3.54 &  33 &    204$\pm$  16 &  12.4 &   6.2$\pm$  0.8 \\
SWJ090946+415713 & 137.44232 &  41.95386 &        12328 &  1.23 &  46 &    133$\pm$  12 &  10.9 &  26.3$\pm$  3.6 \\
SWJ092619-090546 & 141.57982 &  -9.09634 &        46474 &  3.78 &  51 &    251$\pm$  18 &  13.8 &  14.1$\pm$  1.9 \\
SWJ092729+301048 & 141.87399 &  30.18001 &        96630 &  1.72 &  34 &    692$\pm$  28 &  24.4 &  17.7$\pm$  1.9 \\
SWJ092650+301345 & 141.70862 &  30.22919 &       157970 &  1.72 &  23 &    254$\pm$  18 &  13.7 &   4.0$\pm$  0.5 \\
SWJ092719+301342 & 141.83119 &  30.22845 &       152365 &  1.72 &  28 &    312$\pm$  20 &  14.9 &   5.0$\pm$  0.6 \\
SWJ093045+165931 & 142.68851 &  16.99206 &       100039 &  3.59 &  39 &    160$\pm$  16 &   9.9 &   4.1$\pm$  0.6 \\
SWJ093749+153540 & 144.45514 &  15.59446 &        84892 &  3.20 &  21 &    107$\pm$  11 &   9.1 &   3.2$\pm$  0.5 \\
SWJ094816-131644 & 147.07079 & -13.27913 &       165422 &  4.01 &  29 &    279$\pm$  19 &  14.4 &   4.4$\pm$  0.5 \\
SWJ101341+430655 & 153.42323 &  43.11535 &       135197 &  1.39 &  23 &    170$\pm$  14 &  11.4 &   3.1$\pm$  0.4 \\
SWJ105946+534809 & 164.94243 &  53.80263 &        18571 &  0.84 &  54 &    177$\pm$  14 &  12.2 &  23.0$\pm$  3.0 \\
SWJ115811+452906 & 179.54733 &  45.48522 &       107984 &  1.27 &  33 &    183$\pm$  16 &  11.4 &   4.1$\pm$  0.6 \\
SWJ123620+285905 & 189.08401 &  28.98488 &        34342 &  1.55 &  38 &    140$\pm$  14 &   9.7 &  10.0$\pm$  1.5 \\
SWJ124312+170451 & 190.80287 &  17.08105 &        87595 &  1.70 &  21 &    127$\pm$  12 &  10.1 &   3.6$\pm$  0.5 \\
SWJ131300+080259 & 198.25122 &   8.04983 &        70650 &  2.10 &  29 &    209$\pm$  16 &  12.5 &   7.4$\pm$  1.0 \\
SWJ131522+164145 & 198.84331 &  16.69597 &        15244 &  1.69 &  48 &    169$\pm$  14 &  11.8 &  27.4$\pm$  3.6 \\
SWJ133055+420017 & 202.73051 &  42.00475 &       160751 &  0.96 &  47 &    755$\pm$  31 &  24.2 &  11.3$\pm$  1.2 \\
SWJ133051+420647 & 202.71504 &  42.11332 &       159507 &  0.96 &  26 &    133$\pm$  14 &   9.4 &   2.0$\pm$  0.3 \\
SWJ140637+274349 & 211.65532 &  27.73030 &        72908 &  1.65 &  47 &    935$\pm$  32 &  28.8 &  31.5$\pm$  3.4 \\
SWJ140639+273546 & 211.66479 &  27.59635 &        66926 &  1.65 &  47 &    457$\pm$  23 &  19.2 &  16.8$\pm$  2.0 \\
SWJ140728+274917 & 211.86679 &  27.82161 &        63067 &  1.65 &  54 &    324$\pm$  21 &  15.0 &  12.7$\pm$  1.7 \\
SWJ143646+275157 & 219.19327 &  27.86590 &       209771 &  1.86 &  19 &    134$\pm$  14 &   9.3 &   1.6$\pm$  0.2 \\
SWJ151324+305738 & 228.35167 &  30.96066 &       344786 &  1.79 &  14 &    116$\pm$  13 &   8.8 &   0.8$\pm$  0.1 \\
SWJ155742+353023 & 239.42760 &  35.50648 &       141816 &  2.05 &  54 &  11320$\pm$ 107 & 105.2 & 198.3$\pm$ 20.8 \\
SWJ164956+313021 & 252.48631 &  31.50598 &       102390 &  2.45 &  21 &    157$\pm$  13 &  11.5 &   3.9$\pm$  0.5 \\
SWJ173721+461834 & 264.33838 &  46.30944 &       132394 &  2.28 &  37 &    475$\pm$  24 &  19.1 &   9.0$\pm$  1.1 \\
SWJ173932+272055 & 264.88513 &  27.34871 &       134973 &  3.91 &  28 &    290$\pm$  19 &  15.0 &   5.6$\pm$  0.7 \\
SWJ175640+332928 & 269.16901 &  33.49131 &        34848 &  3.56 &  32 &    138$\pm$  12 &  10.7 &  10.2$\pm$  1.5 \\
SWJ181053+581527 & 272.72318 &  58.25763 &        10240 &  3.16 &  54 &    119$\pm$  11 &  10.0 &  29.8$\pm$  4.4 \\
SWJ194004+782419 & 295.01709 &  78.40552 &        91021 &  5.78 &  37 &    158$\pm$  15 &  10.2 &   4.8$\pm$  0.7 \\
SWJ203723-440141 & 309.34961 & -44.02815 &        62280 &  2.98 &  35 &    110$\pm$  13 &   8.2 &   4.5$\pm$  0.7 \\
SWJ215507+164725 & 328.78226 &  16.79029 &       316256 &  5.66 &  16 &    144$\pm$  14 &   9.6 &   1.2$\pm$  0.2 \\
SWJ215354+165348 & 328.47513 &  16.89677 &        57331 &  5.66 &  48 &    337$\pm$  21 &  15.6 &  16.0$\pm$  2.0 \\
SWJ222600-571248 & 336.50238 & -57.21351 &        29306 &  1.83 &  54 &    398$\pm$  21 &  18.2 &  33.6$\pm$  4.9 \\
SWJ222443-022031 & 336.18286 &  -2.34217 &       206336 &  4.26 &  22 &    154$\pm$  15 &  10.2 &   2.0$\pm$  0.3 \\
SWJ222516-020827 & 336.31879 &  -2.14101 &       162526 &  4.26 &  25 &    156$\pm$  15 &  10.3 &   2.5$\pm$  0.4 \\
SWJ222437-022230 & 336.15738 &  -2.37527 &       118954 &  4.26 &  29 &    170$\pm$  15 &  10.9 &   3.8$\pm$  0.5 \\
SWJ222917-110106 & 337.32349 & -11.01850 &       179677 &  4.09 &  20 &    216$\pm$  16 &  13.1 &   3.2$\pm$  0.4 \\
SWJ222953+194354 & 337.47308 &  19.73191 &       134665 &  4.41 &  52 &    679$\pm$  30 &  22.3 &  13.3$\pm$  1.6 \\
SWJ224207+233354 & 340.53061 &  23.56503 &        48472 &  4.77 &  39 &    237$\pm$  17 &  13.8 &  13.1$\pm$  1.7 \\
SWJ230754-681505 & 346.97769 & -68.25166 &        82185 &  2.75 &  23 &    147$\pm$  13 &  10.6 &   4.5$\pm$  0.6 \\
SWJ230650-680401 & 346.71140 & -68.06695 &        82160 &  2.75 &  54 &    596$\pm$  28 &  20.7 &  18.4$\pm$  2.3 \\
SWJ232248+054809 & 350.70193 &   5.80263 &       200154 &  5.04 &  39 &   1549$\pm$  42 &  36.1 &  20.8$\pm$  2.2 \\
SWJ232345-313048 & 350.93924 & -31.51346 &        99340 &  1.16 &  34 &    338$\pm$  20 &  16.4 &   8.3$\pm$  1.0 \\
SWJ233518-662139 & 353.82767 & -66.36105 &       111461 &  2.82 &  30 &    226$\pm$  17 &  13.0 &   5.1$\pm$  0.7 \\
SWJ233617-313626 & 354.07169 & -31.60723 &        43325 &  1.22 &  54 &   1400$\pm$  39 &  35.1 &  78.5$\pm$  9.1 \\
\hline
\end{longtable}
\tablefoot{ (1) SXCS Id; (2),(3) Coordinates of the X--ray
  centroid; (4) Exposure time corrected for vignetting; (5) Galactic
  absorption; (6) Extraction radius $R_{ext}$ in arcsec; (7) Soft net counts
  within $R_{ext}$; \;\;\;\;\;\;\;\;\;\;\;\;\;\;\;\;\;\; (8) S/N; (9) Soft band flux corrected for Galactic
  absorption.}
\end{landscape}
}

\longtabL{3}{
\begin{landscape}
\begin{center}
\begin{table*}
\centering
\caption{List of counterparts of SXCS sources.}
\label{counterpart}
\begin{tabular}{@{}lcccccccc@{}}\hline
\smallskip
\bf Name   & \bf   Cluster & \bf  Distance (') & \bf  redshift &\bf   Galaxy   &  \bf  Distance (') & \bf  redshift  &  \bf  X--ray & \bf  Distance (')  \\ 
\hline
 \hline
SWJ005500-385226 & EDCC        &  0.952  &   -         &  LCRS    &  0.066     &  0.164127   &   -     &  -      \\
SWJ021747-500322 &    -        &    -    &   -         &   -      &    -       &   -         & 1AXG    &  0.571  \\
SWJ023924-250504 &    -        &  -      &   -         &  2DFGRS  &  0.109     &  0.1737     &   -     &  -      \\
SWJ035259-004342 & WHL         &  0.102  & 0.3251 (ph) &  SDSS    &  0.09      &  0.334925   &   -     &  -      \\
SWJ084749+133141 & WHL         &  0.065  & 0.363  (ph) &  SDSS    &  0.061     &  0.348628   &    -    &  -      \\
SWJ090946+415713 & WHL         &  0.128  & 0.14855 (ph)&     -    &    -       &    -        &    -    &  -      \\
SWJ092729+301048 & WHL         &  0.072  & 0.365 (ph)  &  SDSS    &  0.073     &  0.293 (ph) &  1AXG   &  0.35   \\
SWJ092719+301342 & WHL         &  0.223  & 0.29975 (ph)&    -     &    -       &    -        &  -      &  -      \\
SWJ093045+165931 &     -       &    -    &   -         & 2MASX    &  0.053     &  0.177278   &  -      &  -      \\
SWJ093749+153540 & AMF         &  0.0363 & 0.2939 (ph) &  -       &       -    &     -       &   -     &  -      \\
SWJ094816-131644 &     -       &    -    &   -         &    -     &     -      & -           &  CXO    & 0.107  \\
SWJ101341+430655 &     -       &    -    &   -         & SDSS     &  0.043     &  0.448860   &  -      &  -      \\
SWJ105946+534809 & SDSS-C4-DR3 &  0.065  & 0.072 (ph)  & MCG      &  0.066     &  0.071136   &  -      &  -      \\  
SWJ115811+452906 & AMF         &  0.558  & 0.4051 (ph) &    -     &     -      &     -       &  -      &  -      \\
SWJ123620+285905 & AMF         &  0.272  & 0.2305 (ph) & 2MASX    &  0.053     &  0.228573   &  -      &  -      \\
SWJ124312+170451 & NSCS        &  0.73   & 0.1424 (ph) &  -       &   -        &     -       &  -      &  -      \\
SWJ131300+080259 & WHL         &  0.076  & 0.5598 (ph) &  -       &    -       &   -         &  -      &  -      \\
SWJ131522+164145 &     -       &    -    &   -         & -        &  -         & -           &  1WGA   & 0.558  \\
SWJ133055+420017 &     -       &    -    &   -         & 2MASX    &  0.068     &  0.061154   &   -     &  -      \\
SWJ140637+274349 &     -       &    -    &   -         &  -       &     -      & -           &  1WGA  & 1.083   \\
SWJ140639+273546 & GMBCG       &  0.211  &  0.243 (ph) &  -       &     -      &     -       &   -     &  -      \\
SWJ140728+274917 & Abell       & 0.554   &  0.17244    &  -       &     -      &     -       &  1WGA   & 0.796   \\ 
SWJ143646+275157 & AMF         &  0.706  & 0.2648 (ph) &     -    &     -      &       -     &     -   &  -      \\
SWJ151324+305738 & CGCG        & 0.092   &   0.0717    &          & -          &  -          &  -      &  -      \\
SWJ155742+353023 & MaxBCG      & 0.547   & 0.1549      & 2MASX    & 0.117      & 0.158877    & RXJ/WGA & 0.095/0.21 \\
SWJ215507+164725 &     -       &    -    &   -         &          & -          &   -         &  CXO    & 0.062   \\
SWJ222600-571248 & Abell       &  0.268  & 0.13        &   -      &     -      &       -     &   -     &  -      \\
SWJ222516-020827 &     -       &    -    &   -         & -        & -          &    -        &  1WGA   & 1.393   \\
SWJ232248+054809 & NSCS        & 0.914   & 0.45 (ph)   &   -      &    -       &   -         &  -      &  -      \\
SWJ233617-313626 & Abell       & 0.63    & 0.0623 (ph) &   -      &    -       &    -        &  -      &  -      \\  
\smallskip
\end{tabular}
\tablefoot{ (1) SXCS Id; (2) Cluster Optical catalog (3); distance
  from the X--ray centroid in arcmin; (4) cluster redshift; (5) Galaxy
  Optical catalog; (6) distance from the X--ray centroid in arcmin; (7) galaxy
  redshift; (8) X--ray catalog; (9) distance from the X--ray centroid in arcmin.  }
\end{table*}
\end{center}
\end{landscape}
}

\end{document}